\documentclass[prb,twocolumn,showpacs,amsmath,amssymb]{revtex4}
\usepackage{graphicx}
\usepackage{dcolumn} 
\usepackage{bm}      
\newcommand{\be}{\begin{equation}}
\newcommand{\ee}{\end{equation}}
\newcommand{\bea}{\begin{eqnarray}}
\newcommand{\eea}{\end{eqnarray}}
\newcommand{\bw}{\begin{widetext}}
\newcommand{\ew}{\end{widetext}}
\newcommand{\mm}{\mathrm}

\newcommand{\ud}{\mm{d}}
\newcommand{\half}{\frac{1}{2}}

\newcommand{\nn}{\nonumber\\}
\newcommand{\bi}{\begin{itemize}}
\newcommand{\ei}{\end{itemize}}
\newcommand{\etal}{\textit{et al. }}

\begin{document}

  \title{Determination of intrinsic switching field distributions in perpendicular recording media:
    numerical study of the $\Delta H(M, \Delta M)$ method}
  \author{Yang Liu}
  \author{Karin A.\ Dahmen}
  \affiliation{Department of Physics, University of Illinois at
    Urbana-Champaign, Urbana, IL 61801, USA}
  \author{A. Berger}
  \affiliation{San Jose Research Center, Hitachi Global Storage Technologies,
    San Jose, CA 95135, USA}

  \date{\today}
  \begin{abstract}
    We present a numerical study of the $\Delta H(M,\Delta M)$ method and its
    ability to accurately determine intrinsic switching field
    distributions in interacting granular magnetic materials such as
    perpendicular recording media. In particular, we study how this
    methodology fails for large ferromagnetic inter-granular interactions, at
    which point the associated strongly correlated magnetization reversal
    cannot be properly represented by the mean-field approximation, upon which
    the $\Delta H(M,\Delta M)$ method is based. In this study, we use a
    2-dimensional array of symmetric hysterons that have an intrinsic
    switching field distribution of standard deviation $\sigma$ and ferromagnetic
    nearest-neighbor interactions $J$. We find the $\Delta H(M,\Delta M)$ method
    to be very accurate for small $J/\sigma$ values,
    while substantial errors develop once the effective exchange field becomes
    comparable with $\sigma$, corroborating earlier results from micromagnetic
    simulations. We furthermore demonstrate that this failure is correlated
    with deviations from data set redundancy, which is a key property of the
    mean-field approximation. Thus, the $\Delta H(M,\Delta M)$ method fails in
    a well defined and quantifiable manner that can be easily assessed from
    the data sets alone.
  \end{abstract}


  \maketitle

  \section{Introduction}
  One of the key challenges in advancing the nanotechnology of magnetic
  recording is the optimization of recording media and its physical
  properties~\cite{Plumer-01}. This challenge is particularly demanding
  because magnetic recording is by its nature a local process. Thus, it is not
  so much the average physical properties that are crucial, but the
  distributions of such properties that determine continued technological
  advancement and success~\cite{Plumer-01}. In general, it
  is important to devise recording media structures that have very homogeneous
  properties on recording relevant length scales, so that a position independent
  physical description of all properties and magnetization processes is
  appropriate. However, this can only be achieved to a limited degree and it
  is therefore essential to have exact knowledge of the corresponding parameter
  distributions. One of the most crucial properties is the
  intrinsic switching-field distribution $D(H_\mm{S})$ of the media grains because it
  defines the recording quality of a media layer in both magnetic stability and
  the achievable recording density~\cite{Shimizu-03}. Hereby, one has to realize that it is
  not the macroscopic switching field distribution $D_\mm{m}(H_\mm{S})$ in a
  uniformly applied field that is relevant, but the local distribution
  $D(H_\mm{S})$ of switching fields in a recording process, which takes place
  in a narrowly defined field geometry. The difference between
  $D_\mm{m}(H_\mm{S})$ and $D(H_\mm{S})$ is caused by the inter-granular
  interactions between the media grains. In particular, for perpendicular
  media the dipolar interaction is large and dominates the behavior of
  hysteresis loops $M(H)$. Thus, the knowledge of $D_\mm{m}(H_\mm{S})$, which
  can be derived from the slope of $M(H)$, is insufficient for recording
  performance predictions.

  Over the years, several methodologies have been developed to determine
  $D(H_\mm{S})$ with varying success~\cite{Tagawa-91,Pike-99,
  Veerdonk-02,Veerdonk-03,Berger-05, Berger-06,Winklhofer-06}. Most of these
  methods apply a measurement scheme, in which magnetization reversal of media
  grains is probed starting from different magnetization states to allow for a
  variation of the grain-to-grain interaction under measurement
  conditions. Such procedures should then in principle and under certain
  limiting conditions allow for a separation of the intrinsic switching field
  distribution and the inter-granular interactions.

  In this paper, we study the reliability of the recently developed
  $\Delta H(M,\Delta M)$ method, which has been used
  successfully in analyzing progress in recording media
  fabrication~\cite{Berger-05, Berger-06}. The method itself is a
  generalization of an earlier measurement technique, the $\Delta H_\mm{C}$
  method~\cite{Tagawa-91}, but has the advantage that
  it allows the determination of the entire $D(H_\mm{S})$ distribution and not
  just a single characteristic parameter. Furthermore, it enables
  oversampling, which makes consistency checks feasible and gives one the
  opportunity to quantify the confidence level of measurement results. In
  contrast to the also quite robust method developed by van de Veerdonk
  \etal~\cite{Veerdonk-02,Veerdonk-03}, it does not rely on a specific
  distribution form 
  or the rather limiting assumption that interactions can be removed from the
  problem by a simple de-shearing of the major hysteresis loop. Recent data
  indicate that this very assumption appears to be an overly simplistic view
  of inter-granular interactions~\cite{Berger-06}. Another method that has
  recently gained certain popularity is the FORC-method~\cite{Pike-99,
  Winklhofer-06}, which is very similar to the measurement of Preisach
  distributions~\cite{Mayergoyz-86}. However, this technique cannot
  really be compared to the previously mentioned methods, because it
  represents for the most part a data transformation tool and does not appear
  to allow a self-consistent way of extracting microscopic information such as
  $D(H_\mm{s})$, because all interactions are removed from the model in an ad
  hoc fashion simply by definition~\cite{Pike-99, Winklhofer-06}.

  Here, we present a numerical study of the $\Delta H(M,\Delta M)$ method at
  zero temperature. We show that even though this method approximates the
  inter-granular interactions on the mean-field level, it can predict its own
  reliability correctly. The paper is organized as follows. In
  Sec.~\ref{sec:method}, we give a brief introduction to the $\Delta
  H(M,\Delta M)$ method. In Sec.~\ref{sec:model}, we discuss the model used to
  numerically calculate the magnetization curves $M(H)$. In
  Sec.~\ref{sec:reliability}, we introduce reliability measures to quantify
  the failure of the $\Delta H(M,\Delta M)$ method. In Sec.~\ref{sec:results},
  we present and analyze our numerical results and a summary is given in
  Sec.~\ref{sec:sum}.

  \section{\label{sec:method} $\Delta H (M,\Delta M)$ Method}

\begin{figure}[t]
    \includegraphics[width=0.5\textwidth]{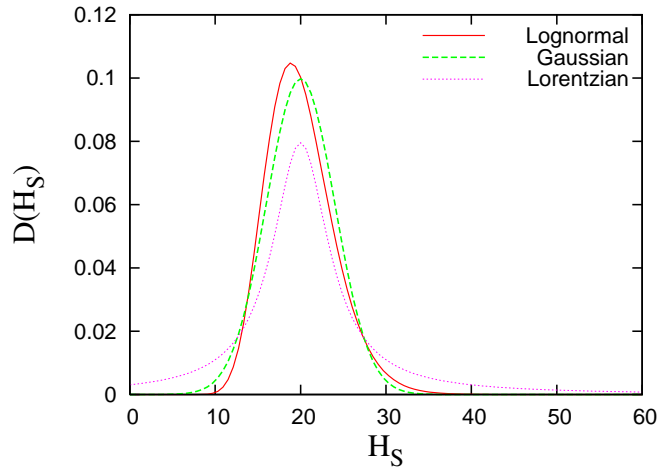}
    \caption{\label{fig:distributions} Gaussian, Lorentzian and Lognormal
      $D(H_\mm{S})$ distributions with the same disorder parameter $\sigma$. For
      both Gaussian and Lognormal distribution, the standard deviation is set to
      be 4.0 and the mean value 20.0. For the Lorentzian distribution, we
      set the width at half-maximum to be 4.0 and the center of the distribution
      to be 20.0. }
  \end{figure}

  The $\Delta H (M,\Delta M)$ method assumes that the effective field at each
  grain can be written as $H_\mm{eff}=H(M)+H_\mm{int}(M)$ with $H(M)$ being the
  external field and $H_\mm{int}(M)$ the volume-averaged interaction field at
  magnetization $M$. Assuming that $M$ is normalized to its saturation value
  $M_\mm{S}$, the upper branch of the major hysteresis loop driven by the
  major loop external field $H$, and the recoil curve starting at
  $M_\mm{rev}=1-\Delta M$ driven by the recoil curve external field
  $H_\mm{r}$ are given as:

  \begin{subequations}
    \begin{align}
      M &= 1 - 2 \int_{-\infty}^{-[H(M)+H_\mm{int}(M)]} D(H_\mm{S}) \, \ud H_\mm{S}\\
      M &= 1-\Delta M - 2 \int_{-\infty}^{-[H_\mm{r}(M)+H_\mm{int}(M)]}
      D(H_\mm{S}) \, \ud H_\mm{S}
    \end{align}
  \end{subequations} respectively, i.e. field integrals over the intrinsic
  switching field distribution $D(H_\mm{S})$. Defining an integral function
  $I(x) = \int_{-\infty}^{x} D(H_\mm{S}) \, \ud H_\mm{S}$, one finds

  \begin{subequations}
    \begin{align}
      H(M) &= -I^{-1}\Big( \frac{1-M}{2}\Big) - H_\mm{int}(M) \\
      H_\mm{r}(M) &= -I^{-1}\Big( \frac{1-M-\Delta M}{2}\Big) - H_\mm{int}(M)
    \end{align}
  \end{subequations} with $I^{-1}$ being the inverse of the integral
  function. Therefore, one can derive

  \bea \Delta H(M,\Delta M)
  &\equiv& H_\mm{r}(M)-H(M) \nn
  &=& I^{-1}\Big(\frac{1-M}{2}\Big) - I^{-1}\Big( \frac{1-M-\Delta M}{2}\Big) \nn
  \label{eq:DeltaH}\eea as a closed functional form for the field axis
  distance $\Delta H$ between major and recoil curves, a quantity that is
  illustrated in Fig.~\ref{fig:redundancy_MH}. Important is hereby that
  within the framework of this approach $\Delta H$ is independent from the
  grain interaction, which allows for a direct experimental access to
  $D(H_\mm{S})$. For certain parameterized distribution functions, one can
  derive analytic expressions for $\Delta H$. For example, for the Gaussian,
  Lorentzian and Lognormal $D(H_\mm{S})$ distributions (see
  Fig.~\ref{fig:distributions}),

  \begin{subequations}
    \begin{align}
      D_\mm{G}(H_\mm{S})&= \frac{1}{\sqrt{2\pi} \sigma}
      \exp\left[-\frac{(H_\mm{S}-h_0)^2}{2\sigma^2} \right] \label{eq:Gaussian}\\
      D_\mm{L}(H_\mm{S})&= \frac{2w}{\pi}\frac{1}{w^2+4(H_\mm{S}-h_0)^2} \label{eq:Lorentzian}\\
      D_\mm{LN}(H_\mm{S})&= \frac{1}{\sqrt{2\pi} \tilde{\sigma} H_\mm{S}}
      \exp\left[-\frac{(\log H_\mm{S}-\tilde{\mu})^2}{2\tilde{\sigma}^2} \right] \label{eq:Lognormal}
    \end{align}
  \end{subequations} we find

  \begin{subequations}
    \begin{align}
      \Delta H_\mm{G}(M,\Delta M) &=\sqrt{2}  \sigma  \left( \mm{erf}^{-1}(M+\Delta
      M) - \mm{erf}^{-1}(M) \right) \label{eq:DeltaH-Gaussian}\\
      \Delta H_\mm{L}(M,\Delta M) &=\frac{w}{2} \left\{ \tan[\frac{\pi}{2}(M+\Delta
	M)] - \tan[ \frac{\pi}{2}M] \right\}\\
      \Delta H_\mm{LN}(M,\Delta M) &= \exp\big[\tilde{\mu}-\sqrt{2} \tilde{\sigma}
	\mm{erf}^{-1}(M)\big] \nn & \ \ -\exp\big[\tilde{\mu}-\sqrt{2} \tilde{\sigma}
	\mm{erf}^{-1}(M+\Delta M)\big] \label{eq:distributions}
    \end{align}
  \end{subequations}

  Results for the Gaussian and Lorentzian distributions were reported
  previously~\cite{Berger-05, Berger-06}. Here, we introduce the disorder
  parameter $\sigma$ for a general $D(H_\mm{S})$ distribution. $\sigma$ is
  defined to be the standard deviation of a given distribution,
  such as Gaussian and Lognormal. However, for the Lorentzian distribution
  neither variance nor higher moments are defined, so that we need to quantify
  the disorder level in another form. For the Lorentzian, we define the
  disorder parameter to be the distribution width at the half-maximum and the
  mean to be the center of the distribution. For the
  distributions given by Eq.~\ref{eq:Gaussian},~\ref{eq:Lorentzian} and
  ~\ref{eq:Lognormal},  we then have the disorder parameter:
  $\sigma^\mm{G} = \sigma$, $\sigma^\mm{L}  = w$, $\sigma^\mm{LN} =
  e^{\tilde{\mu}+\tilde{\sigma}^2/2} (e^{\tilde{\sigma}^2}-1)^{1/2}$  and the
  mean value $h_0^\mm{G} = h_0$, $h_0^\mm{L} = h_0$,
  $h_0^\mm{LN} = e^{\tilde{\mu}+\tilde{\sigma}^2/2}$.

  By making a least-squares fit of the $\Delta H(M,\Delta M)$ curves to the above formulas, one can
  extract the key features of $D(H_\mm{S})$. Note that both $\Delta
  H_\mm{G}(M,\Delta M)$ and $\Delta H_\mm{L}(M,\Delta M)$ have no $h_0$
  dependence. But for the Lognormal distribution, $\Delta H_\mm{LN}(M,\Delta
  M)$ depends on both $\tilde{\mu}$ and $\tilde{\sigma}$, and therefore it has
  both $\sigma$ and $h_0$ dependencies.

  It is easy to prove that simply shifting a general distribution will not alter the
  $\Delta H$ data. For this, we consider the case of a general distribution
  $D(H_\mm{S})$: if we shift it towards right by an amount $H_0$, then the
  integral function $\tilde{I}(x)$ of the new distribution
  $\tilde{D}(H_\mm{S})=D(H_\mm{S}-H_0)$ is given by $\tilde{I}(x)=I(x-H_0)$
  and its inverse $\tilde{I}^{-1}(y) = H_0 + I^{-1}(y)$. It is then clear that
  this shift will not change the $\Delta H(M,\Delta M)$ formula at all, since
  the $H_0$ terms will cancel according to Eq.~\ref{eq:DeltaH}.

  Also, we find that both $\Delta H_\mm{G}(M,\Delta M)$ and $\Delta
  H_\mm{L}(M,\Delta M)$ are symmetric with respect to $M=-\Delta M/2$ while
  $\Delta H_\mm{LN}(M,\Delta M)$ is not symmetric. This is consistent with the
  original distribution: Both $D_\mm{G}(H_\mm{S})$ and
  $D_\mm{L}(H_\mm{S})$ are symmetric around $h_0$ while
  $D_\mm{LN}(H_\mm{S})$ is not. Actually, for any $D(H_\mm{S})$ distribution
  being symmetric about $h_0$, one finds that $I(h_0+x) = 1 - I(h_0-x)$ and
  $I^{-1}(1/2+y)=-I^{-1}(1/2-y)$, from which it is easy to prove that $\Delta
  H(M,\Delta M)$ is symmetric about $M=-\Delta M/2$, i.e.

  \be
  \Delta H(-\frac{\Delta M}{2} + M^{\prime},\Delta M) = \Delta H(-\frac{\Delta M}{2} - M^{\prime},\Delta M)
  \ee


  \section{\label{sec:model}Hysteron Model}
\begin{figure}[t]
    \includegraphics[width=0.5\textwidth]{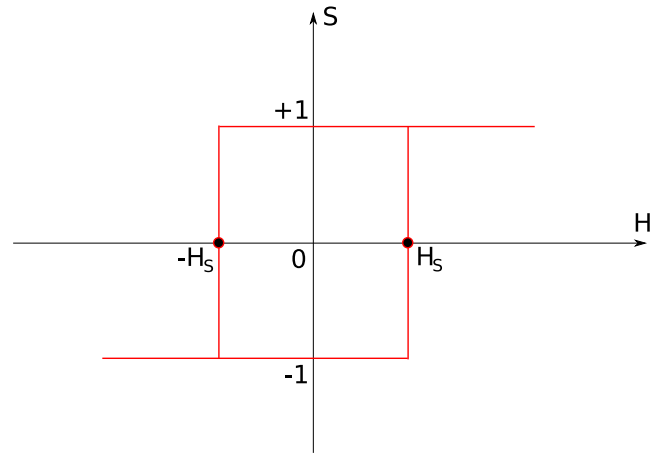}
    \caption{\label{fig:Hysteron} A symmetric hysteron $S$ with intrinsic
      switching field $H_\mm{S}$.}
  \end{figure}

  For our numerical studies, we model each grain as a hysteron, which is the
  simplest mathematical construction for the description of a
  hysteretic system. Each hysteron will generate a
  rectangular hysteresis loop in an applied field $H$ as shown in
  Fig.~\ref{fig:Hysteron}. We assume the transition of each hysteron is
  infinitely sharp and it has no additional field dependence, such as finite 
  susceptibility, for instance. The half-width of the rectangular
  hysteresis loop is referred to as the intrinsic switching field of the
  hysteron, which is a well defined property of each individual hysteron.

  We also assume that the hysteron switching fields are symmetric around zero
  field, i.e. there is no bias. A hysteron with zero bias is called a
  symmetric hysteron and is consistent with the time reversal symmetry that
  ferromagnets exhibit~\cite{Goldenfeld-05}. These hysteron properties can be
  regarded as fairly good representation of perpendicular media grains,
  because they have relatively high magnetic anisotropy and the applied field
  in typical characterization measurements is applied along the easy
  axis~\cite{Berger-05, Berger-06, Tagawa-91, Veerdonk-02, Veerdonk-03,
    Pike-99, Winklhofer-06}.

  We further assume that there is no time structure to the hysteron switch
  itself. This should be an appropriate picture as long as one
  considers field change rates that are much slower than single grain reversal
  times. Given that such reversal times are typically of the order of several
  hundred pico seconds~\cite{Gao-04}, this condition is generally fulfilled in typical
  measurement setups~\cite{Shimatsu-06}.

  In our model the ferromagnetic layer system is then represented by a simple
  square lattice of symmetric hysterons with periodic boundary
  conditions. Note that a square lattice is not necessarily
  a very good approximation of actual media structures, in which grains
  typically have coordination numbers of 5 or 6~\cite{Berger-06}. However, we neglect this
  detail of actual media structures for reasons of simplicity and furthermore
  assume that deviations from a more realistic grain structure are not
  fundamentally altering the overall significance of our study. Under the
  assumptions that hysterons ($S_i=\pm 1$) have an intrinsic switching field distribution
  $D({H_\mm{S}}_i)$ with ${H_\mm{S}}_i>0$, interact ferromagnetically
  with their nearest neighbors with strength $J$ and experience a uniform
  external field $H$, the Hamiltonian of the system can be written as

  \be {\cal H} = - \!\sum_{{<}i,j{>}} J \, S_i S_j - \sum_i \, \Big( H +
  \mm{sgn}(S_i) {H_\mm{S}}_i \Big) \, S_i \label{eq:RSFHM} \ee

  We denote this as the random switching field hysteron model (RSFHM). Note
  that the RSFHM is very similar to the random field Ising model (RFIM):
  \be {\cal H} = - \!\sum_{{<}i,j{>}} J \, s_i s_j - \sum_i \, (H + h_i) \,
  s_i\ee where the spins $s_i = \pm 1$ interact ferromagnetically with their
  nearest neighbors with strength $J$ and experience a uniform external field
  $H$ and a local quenched field $h_i$. A simple mapping: $h_i
  \leftrightarrow \text{sgn}(S_i) {H_\mm{S}}_i$ enables us to
  calculate $M(H)$ curves of symmetric hysterons with the developed
  algorithms used for previous RFIM computational work~\cite{Sethna-93,Kuntz-99}.
  Details of the close relation between these models are described in
  Appendix~\ref{sec:app:mapping}.

  In this paper, we evaluate different types of $D(H_\mm{S})$ to study the
  reliability of the $\Delta H(M,\Delta M)$ method. In our simulation, we set
  the ferromagnetic nearest-neighbor coupling strength $J=1$ and tune the
  disorder parameter $\sigma$. Since $J=1$, tuning $\sigma$ is equivalent to
  tuning $\sigma/J$. In this sense, a small (big) $\sigma$ corresponds to
  strong (weak) nearest-neighbor interactions of hysterons. We calculate the
  $M(H)$ curves, both major hysteresis loop and recoil curves, for system
  sizes up to $1000^2$ and $\sigma$ values from 1.6 to 1000. Note that we are
  ignoring dipolar interactions in our model (Eq.~\ref{eq:RSFHM}), even though
  we know that they are substantial in real structures. But a previous
  micromagnetic study demonstrated that the dipolar interactions can very well
  be treated within the mean-field $\Delta H(M,\Delta M)$ method and do not
  cause any significant precision problems, while exchange interactions
  did~\cite{Berger-06}. Thus, we focus here on the effect of inter-granular
  exchange interactions only, because they represent the much more serious
  problem for the reliability of the $\Delta H(M,\Delta M)$ method. Also, we
  are acutely aware of the fact that the assumption of a uniform exchange
  coupling constant $J$ in Eq.~\ref{eq:RSFHM} is a substantial simplification
  of the problem, if one compares our model to real materials
  structures. However, a lateral varying $J$ would be an extension of the
  present model and will be the topic of future work.

  \section{\label{sec:reliability}Reliability Measures}
  In numerical test simulations of characterization methods, one can introduce
  a simple type of reliability measure, namely the deviation of the retrieved
  fit parameters from the exact solution given by the input parameters of the
  simulation. However, for experimental data such a reliability measure is
  generally not available simply because the materials parameter are not known,
  but are actually what one wants to extract by means of this very
  experimental method. Therefore, in a typical experimental situation one can only guess
  the overall measurement reliability by identifying parameter ranges, in which
  approximate methods are expected to work with a certain level of precision,
  which may or may not be clearly identifiable. One broadly used method to
  identify the reliability of extracted materials parameter is the similarity,
  with which a fit that is based on a certain physical model can reproduce the
  experimental data.

  As we will demonstrate in the following, we encounter an unusually good
  situation for the $\Delta H(M, \Delta M)$ method, in that a consistency test of
  the data set alone can be used to validate the applicability
  of the method and estimate its overall reliability. We will furthermore show
  that this consistency test does not depend on any specific distribution
  shape but just the mean-field approximation in general, i.e. the underlying
  assumption of the $\Delta H(M,\Delta M)$ method.

  In this section, we discuss the reliability measures of the $\Delta
  H(M,\Delta M)$ method for an arbitrary type of $D(H_\mm{S})$. The reliability
  is characterized by two types of measures: (1) conventional quality measures
  for numerical fits such as the percentage difference between the fitting and
  the actual parameter ($P_\mm{d}$) and the square of the multiple correlation
  coefficient ($R^2$); (2) the average deviation from redundancy measure
  ($r$), which represents the above mentioned self-consistency test.

  \subsection{\label{sec:fitquality} Fit Quality}
  In simulations, one knows the input parameters, in our case the input
  $D(H_\mm{S})$ distribution, exactly. Furthermore, we have also derived 
  $\Delta H(M,\Delta M)$ formulas for certain specific $D(H_\mm{S})$
  distributions as shown in Sec.~\ref{sec:method}. Thus, by fitting these
  analytical formulas to the $\Delta H(M,\Delta M)$ curves obtained from our
  2D-RSFHM hysteresis loop simulations, we can get an estimate of the input
  $D(H_\mm{S})$ distribution. Obviously, if the input $D(H_\mm{S})$ is
  recovered by the fitting procedure with high accuracy, then the $\Delta
  H(M,\Delta M)$ method works. To quantify the reliability of the $\Delta
  H(M,\Delta M)$ method, we introduce the following fit quality measures. The
  most important fit quality measure, denoted as $P_\mm{d}$, is the percentage
  difference between the parameters obtained from a least-square fit
  and the actual input parameters into our simulation. It describes how well
  the $\Delta H(M,\Delta M)$ method can indeed retrieve the information
  sought. Focusing on the disorder parameter of $D(H_\mm{S})$ as the most
  crucial fit parameter, we define
  \be P_\mm{d} = (\sigma_\mm{f}-\sigma)/\sigma\ee Here $\sigma_\mm{f}$ is the
  fit value of the $D(H_\mm{S})$ distribution disorder parameter as
  defined in Sec.~\ref{sec:method}, while $\sigma$ is the input value of the
  same parameter.

  Another fit quality measure that can be utilized here is the square of the
  multiple correlation coefficient $R^2$, which measures how successful a fit
  and fit function is in explaining the data~\cite{Draper-66}. It is defined as

  \be R^2 = 1-\frac{\sum_i (Y_i-\hat{Y_i})^2}{\sum_i (Y_i-\bar{Y})^2 }\ee
  $Y_i$ and $\hat{Y_i}$ are hereby the simulation result and the fit function
  value for $\Delta H(M,\Delta M)$ at a given data point $(M_i, \Delta M_i)$,
  respectively. $\bar{Y}$ is the average value of $Y_i$. According to the
  definition of $R^2$, we know that as $R^2$ approaches 
  1, the fit is a better and better representation of the data set. Hereby, it
  is important to emphasize two points in calculating $P_\mm{d}$ and $R^2$. First,
  we are comparing a mean-field theory to numerical simulations that contain
  the complexity of magnetization reversal in its complete detail. Second, we
  are using a finite size grid in our simulation while the analytic theory is
  derived for infinite systems. Naturally, the finite size will affect the
  $R^2$ and $P_\mm{d}$ calculation. Particularly, for data points near the
  beginning or end of the reversal curve, where only a few grains (hysterons)
  are reversed, the analytical theory for infinite hysteron numbers might not
  be accurate at all for the description of a finite system, independent from
  the validity of the mean-field approximation.

  \subsection{\label{sec:r}Deviation From Redundancy}

  \begin{figure}[t]
    \includegraphics[width=0.45\textwidth]   {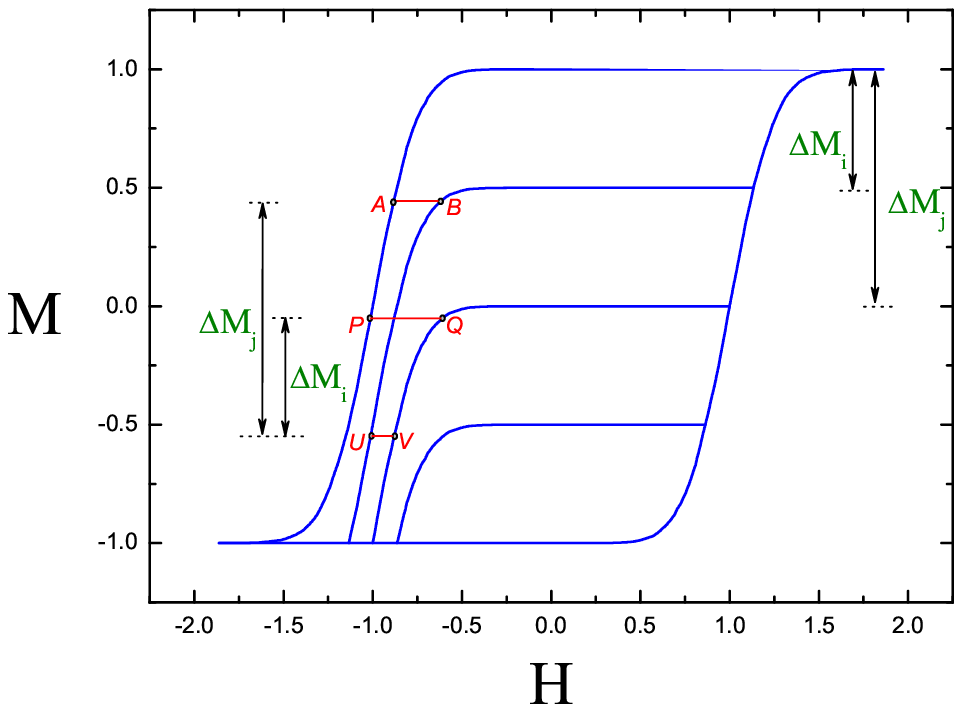}
    \includegraphics[width=0.45\textwidth]   {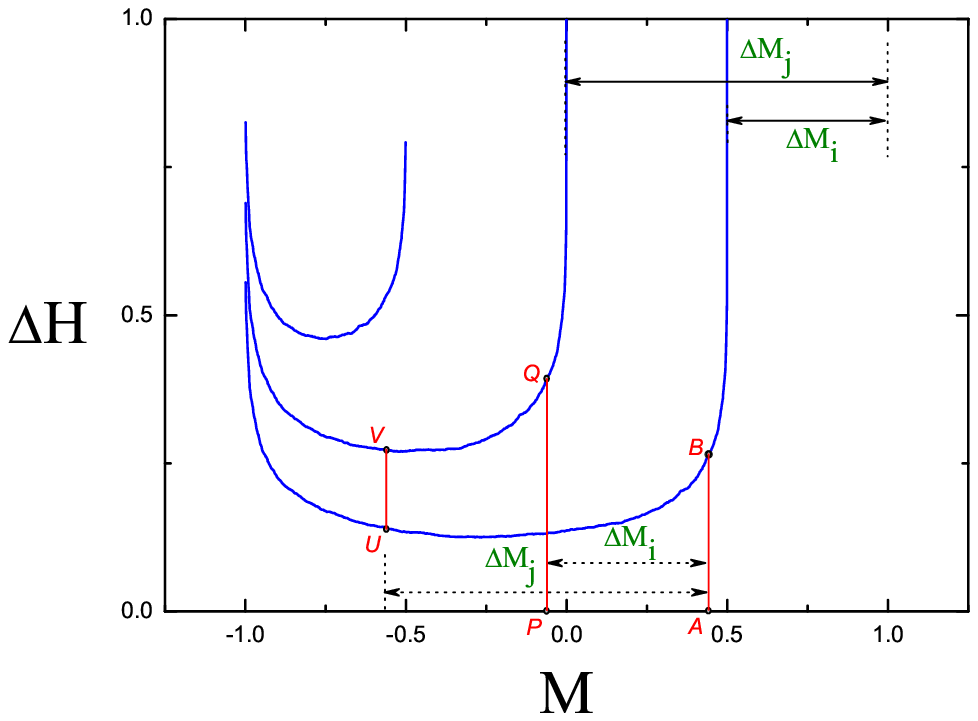}
    \caption{\label{fig:redundancy_MH} (Top)  
      The major hysteresis loop and three recoil curves. Throughout the paper,
      $M$ is normalized to its saturation value $M_\mm{S}$ and $H$ is
      normalized to the coercive field $H_\mm{C}$. The
      first two recoil curves start at $M_\mm{rev}=1-\Delta M_i$ and $1-\Delta M_j$,
      respectively, with $\Delta M_i < \Delta M_j$. For the six points (A and P),
      (B and U), (Q and V) shown in the figure with $M_\mm{A}=M_\mm{B}=M $,
      $M_\mm{P}=M_\mm{Q}=M-\Delta M_{j}+\Delta M_{i}$ and
      $M_\mm{U}=M_\mm{V}=M-\Delta M_{j}$ where $M$ is an arbitrary value
      within $[-1,1-\Delta M_i]$, one can prove the existence of data redundancy,
      i.e. the equality $\left( H_\mm{B} - H_\mm{A} \right) + \left( H_\mm{V} - H_\mm{U}
      \right) = \left( H_\mm{Q} - H_\mm{P} \right)$ within the mean-field
      approximation. (Bottom) The corresponding $\Delta H(M, \Delta M)$ data
      for the three recoil curves.}
  \end{figure}

  As already mentioned above, fit quality measures such as $R^2$ are not
  necessarily foolproof because physical models and corresponding fit
  functions may be used in a regime, for which the underlying theory
  does not apply anymore. In such cases, data fits and extracted
  materials parameters might appear very accurate, while they are not.
  Thus, it would be a tremendous help, if an independent data set
  evaluation would be available that allows a separate measure of
  the suitability of the underlying theory. Specifically here, this
  evaluation should tell us how good an approximation the mean-field
  theory is for any given data set, so that we know to which
  confidence level we can rely on the $\Delta H(M, \Delta M)$ method.

  We find that it is indeed possible to define such a
  quantity. To do so one has to recognize that within the mean-field approximation
  $\Delta H(M, \Delta M)$ data sets contain redundancy. The data set
  redundancy can be seen from Fig.~\ref{fig:redundancy_MH}. For illustration
  purposes, we pick six points: (A and P), (B and U), (Q
  and V) located on the major loop, the $i$-th recoil curve and the $j$-th recoil
  curve, respectively. The $i$-th and $j$-th recoil curve start at
  $M_\mm{rev}=1-\Delta M_i$ and $1-\Delta M_j$, respectively. Without limiting
  the generality of this consideration, we furthermore assume that $\Delta
  M_i < \Delta M_j$ and require that the six points satisfy: $
  M_\mm{A}=M_\mm{B}=M $, $M_\mm{P}=M_\mm{Q}=M-\Delta M_{j}+\Delta M_{i}$ and
  $M_\mm{U}=M_\mm{V}=M-\Delta M_{j}$ where $M$ is an arbitrary value within
  $[-1,1-\Delta M_i]$. We then define

  \begin{subequations}
    \begin{align}
      \Delta H_i (M) =& H_\mm{B} - H_\mm{A} \\
      \Delta H_j (M-\Delta M_{j}+\Delta M_{i}) =& H_\mm{Q} - H_\mm{P} \\
      \Delta H_j (M-\Delta M_{j}) - \Delta H_i (M-\Delta M_{j}) =& H_\mm{V} - H_\mm{U}
    \end{align}
  \end{subequations}

  Within the mean-field approximation, it is easy to prove that
  \be \left( H_\mm{B} - H_\mm{A} \right) + \left( H_\mm{V} - H_\mm{U} \right)
  - \left( H_\mm{Q} - H_\mm{P} \right) = 0 \ee
  as shown in Appendix~\ref{sec:app:proof_redundancy}. More
  generally, one finds that

  \bw

  \be
  \Delta H_i(M) + \Delta H_j(M-\Delta M_j) -\Delta H_i(M-\Delta
  M_j) - \Delta H_j(M-\Delta M_j + \Delta M_i) = 0 \label{eq:redundancy}
  \ee

  This data set redundancy is due to the fact that successive
  recoil curves are not fully independent and contain repeated information. However,
  Eq.~\ref{eq:redundancy} is derived under the assumption of the mean-field
  approximation and is only true if the mean-field approximation is indeed
  fulfilled by the data set. For general data sets this property is not
  conserved. Therefore, we can define an $M$-dependent measure as

  \be r_{ij}(M)=\frac{\Delta H_i(M) + \Delta H_j(M-\Delta M_j) -\Delta H_i(M-\Delta
    M_j) - \Delta H_j(M-\Delta M_j + \Delta M_i)}{\Delta H_i(M) + \Delta
    H_j(M-\Delta M_j) +\Delta H_i(M-\Delta M_j) + \Delta H_j(M-\Delta M_j +
    \Delta M_i)} \label{eq:rij} \ee
  \ew that monitors deviations from the mean-field approximation based upon the
  above redundancy criterion (Eq.~\ref{eq:redundancy}). Eq.~\ref{eq:rij} has an
  $M$ definition range of $[\Delta M_j -1, 1-\Delta M_i]$. For a
  general set of multiple recoil curves, the average deviation from
  redundancy measure can be defined as


  \be  r = \frac{1}{n}\sum_{i,j} \left\langle r^2_{ij}(M)
  \right\rangle^{\half} \ee with $n$ being the total number of all the possible
  $(i,j)$ pairs~\cite{note:rij}. Thus, $r$ is a quantitative measure
  that allows an accurate and independent check of how close or far any
  $\Delta H(M,\Delta M)$ data set is from fulfilling the mean-field approximation.

  Note that in calculating $r$, we are comparing the data set with
  itself. Thus, the validity of the underlying mean-field approximation used
  for all the $\Delta H(M,\Delta M)$ data fits, can be assessed independently
  and from the data set alone.

  An additional benefit in the calculation of $r$ is that the finite size
  inaccuracies at the definition range boundaries for the recoil curves will
  cancel out, at least to some degree. See Eq.~\ref{eq:rij}. For $M=1-\Delta
  M_i$, the finite size inaccuracies in $\Delta H_i(M)$ and $\Delta
  H_j(M-\Delta M_j+\Delta M_i)$ will cancel out. Similarly, for $M=\Delta
  M_j-1$, the finite size inaccuracies in $\Delta H_j(M-\Delta M_j)$ and
  $\Delta H_i(M-\Delta M_j)$ will cancel out. In this sense, the
  deviation from redundancy measure is more robust than the fit quality
  measures.

  \section{\label{sec:results}Simulation Results}
  To show that the $\Delta H(M, \Delta M)$ method fails reproducibly in a
  well-defined manner, we calculate $\Delta H(M, \Delta M)$ data sets for 4
  different distributions. 

  For all these distributions, we tune the disorder parameter $\sigma$ from 1.6
  to 1000, but keep the ratio $h_0/\sigma$ to be a positive constant. This is done to
  avoid any negative switching fields, which would otherwise describe
  non-physical behavior in violation of the second law of
  thermodynamics~\cite{Landau-84}. For a Gaussian, $h_0/\sigma = 5$ is generally big
  enough to avoid any negative $H_\mm{S}$ for system sizes of up to $1000^2$. For a
  Lorentzian, we choose $h_0/\sigma=5\times 10^3$ for systems of size $100^2$ and
  $h_0/\sigma \sim 2\times 10^5$ for systems of size $1000^2$. To avoid
  the long negative tails of the Lorentzian, we can truncate the distribution
  instead of choosing a huge $h_0/\sigma$ ratio, creating a new type of
  distribution, which we refer to as truncated Lorentzian distribution
  $D_\mm{L_t}(H_\mm{S})$ in the following. Further details, including the
  definition of $\sigma$, for this distribution are described in
  Appendix~\ref{sec:app:truncate_Lorentzian}. For this $D_\mm{L_t}(H_\mm{S})$
  distribution, we also choose $h_0/\sigma=5$ in our calculations.
  For the Lognormal distribution, there is by definition no distribution density
  for negative fields. But we still choose $h_0/\sigma=5$ to make it comparable
  with the Gaussian and the truncated Lorentzian distribution.

  \subsection{Comparison with the mean-field approximation}

\begin{figure*}[t]
    \includegraphics[width=2.2in]{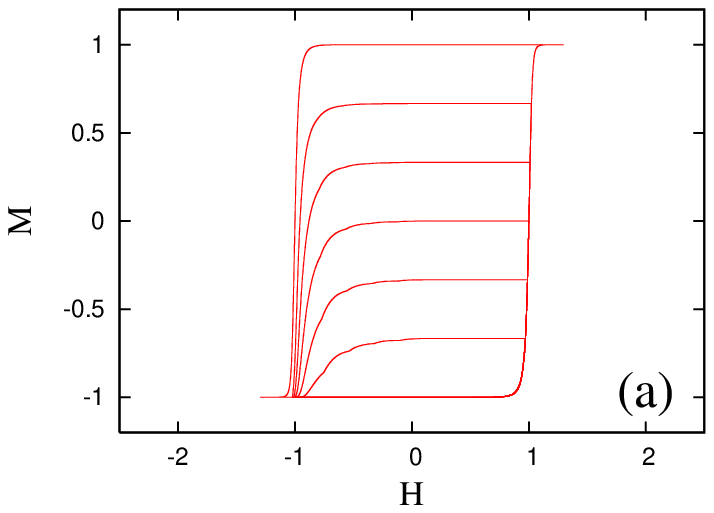}
    \includegraphics[width=2.2in]{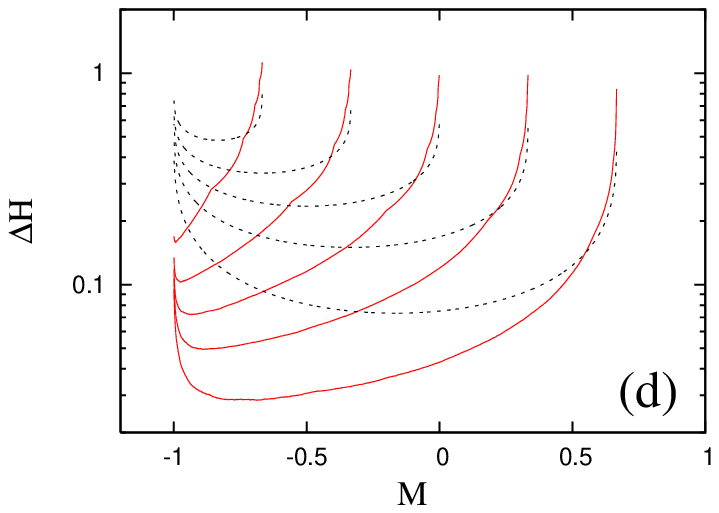}
    \includegraphics[width=2.2in]{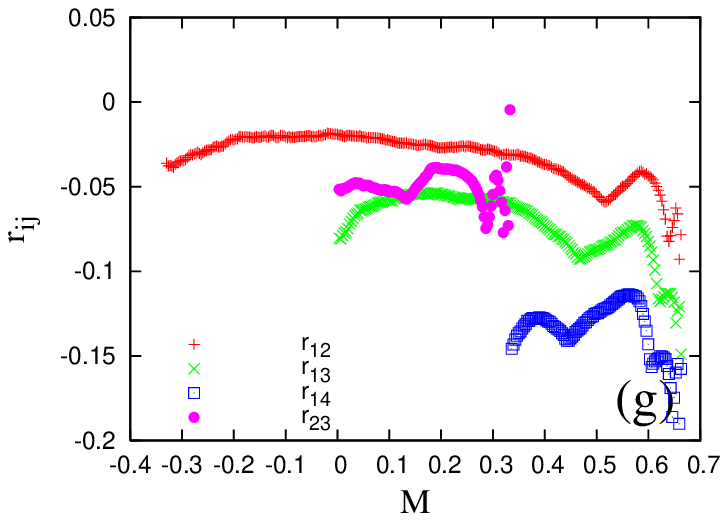}
    \includegraphics[width=2.2in]{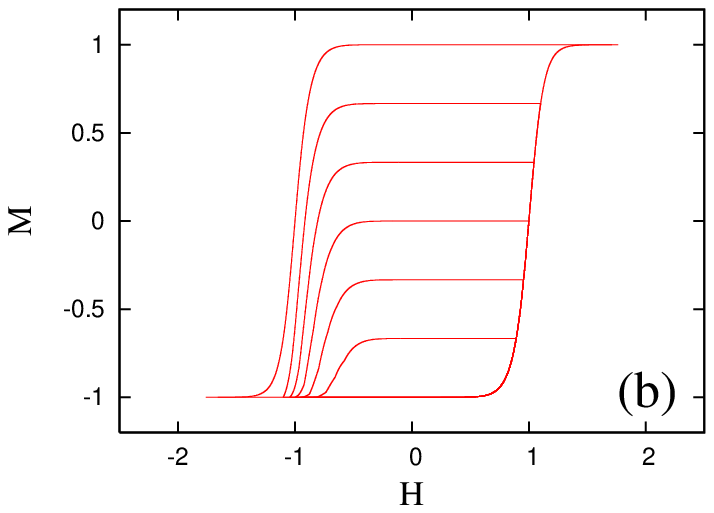}
    \includegraphics[width=2.2in]{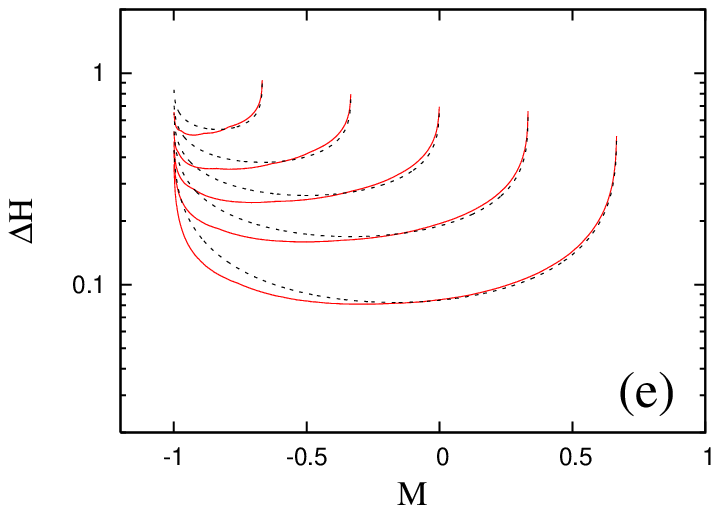}
    \includegraphics[width=2.2in]{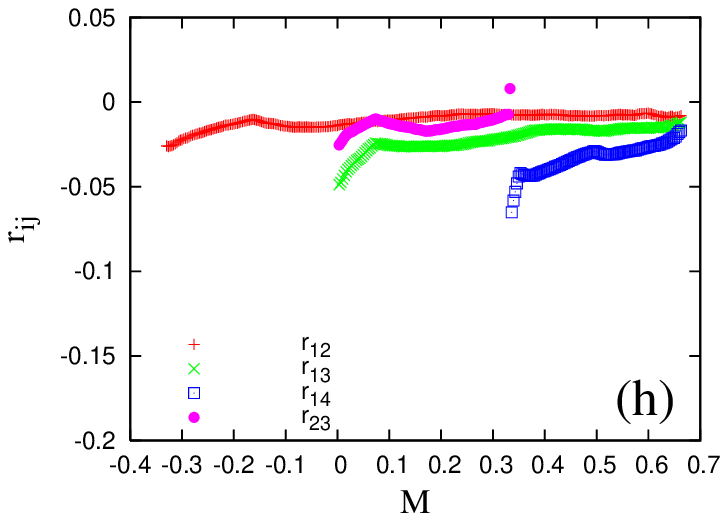}
    \includegraphics[width=2.2in]{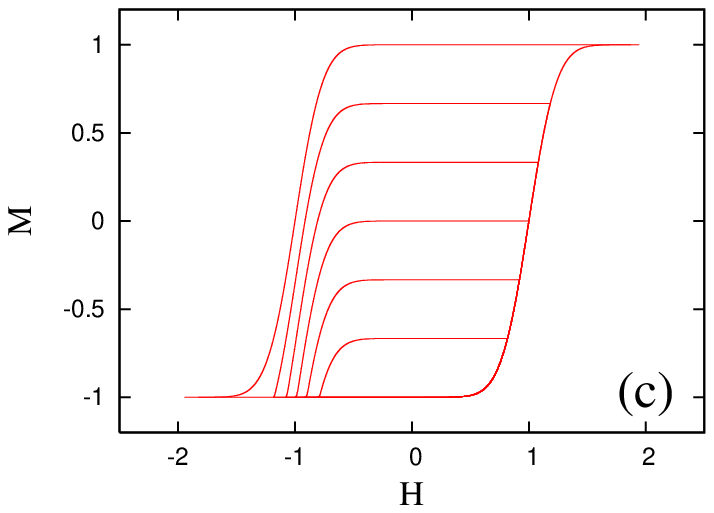}
    \includegraphics[width=2.2in]{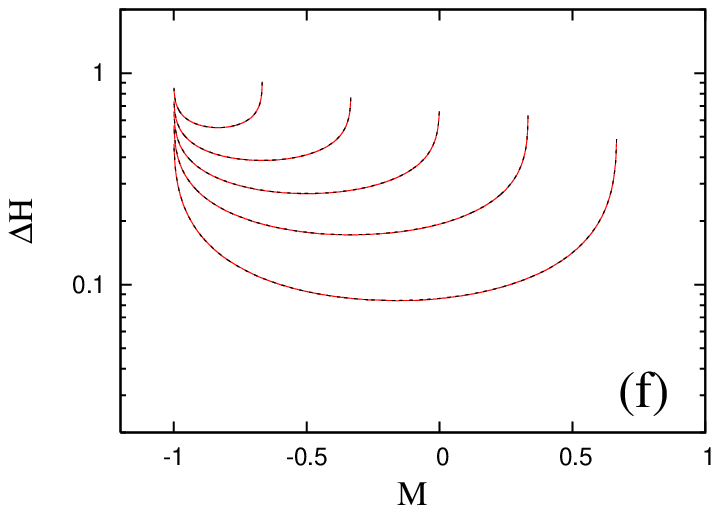}
    \includegraphics[width=2.2in]{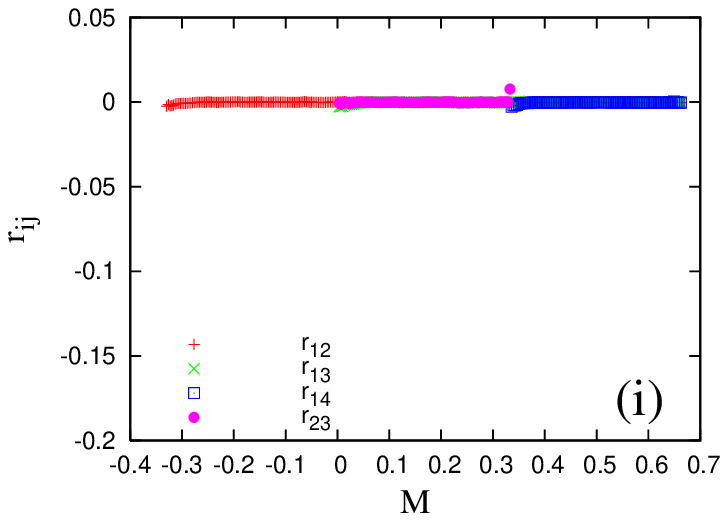}
    \caption{\label{fig:Gaussian} Numerical results using a Gaussian distribution
      RSFHM. Rows: (Top, fig.a,d,g): $\sigma=1.6$. (Middle, fig.b.e.h): $\sigma=5$. (Bottom,
      fig.c,f,i): $\sigma=50$. Columns: (Left, fig.a,b,c): $M(H)$ curves, main loop and 5 recoil
      curves. (Middle, fig.d,e,f): $\Delta H(M, \Delta M)$ curves for the 5 recoil curves: (solid
      lines) numerical result; (dotted lines) mean-field approximation.
      (Right, fig.g,h,i): $M$-dependent deviation from redundancy ($r_{ij}(M)$) for all the
      possible recoil curve pairs.}
  \end{figure*}

\begin{figure*}[t]
    \includegraphics[width=2.2in]{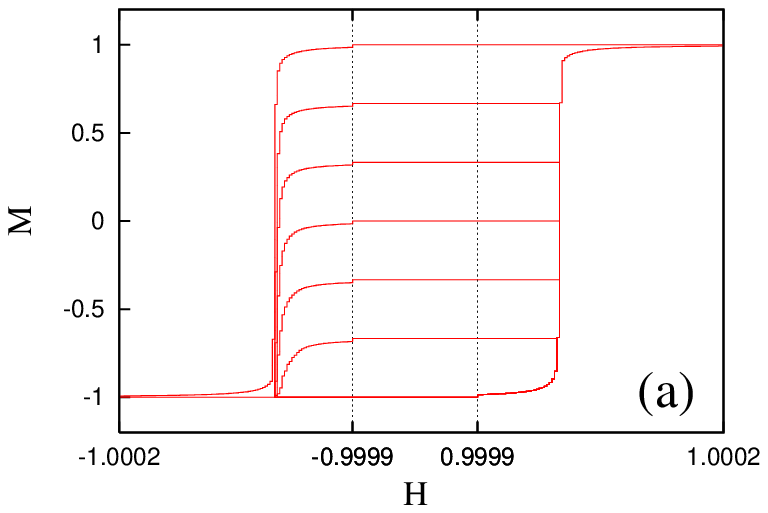}
    \includegraphics[width=2.2in]{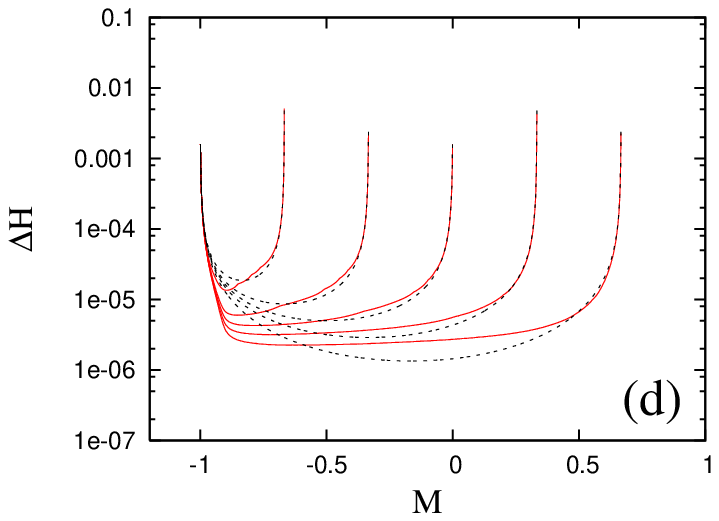}
    \includegraphics[width=2.2in]{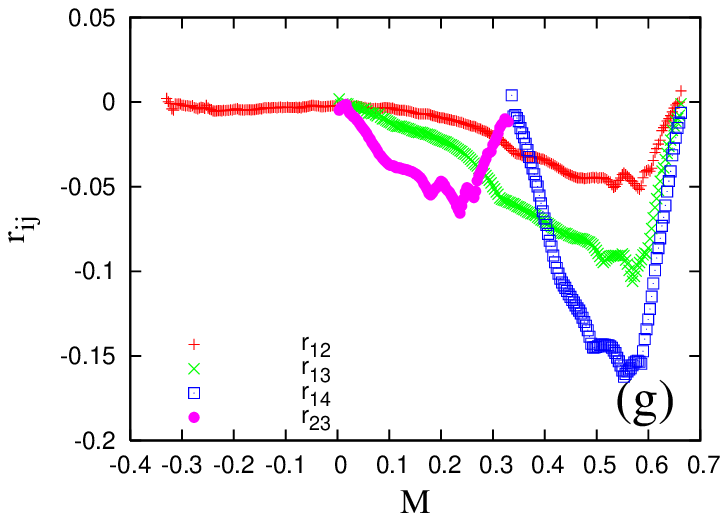}
    \includegraphics[width=2.2in]{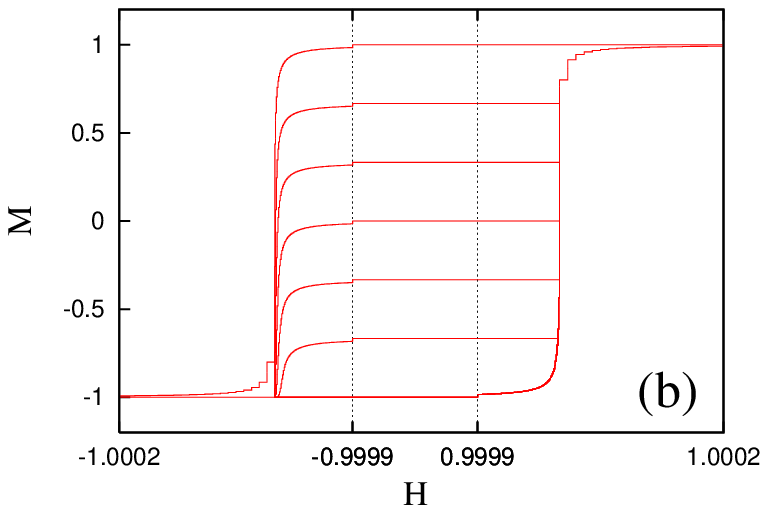}
    \includegraphics[width=2.2in]{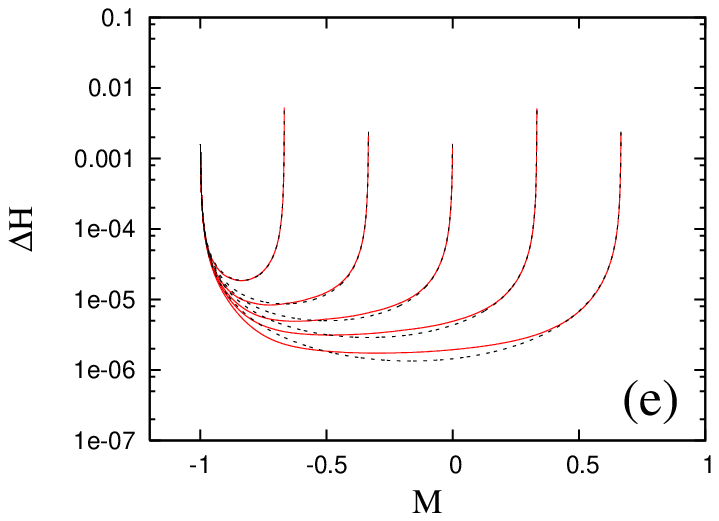}
    \includegraphics[width=2.2in]{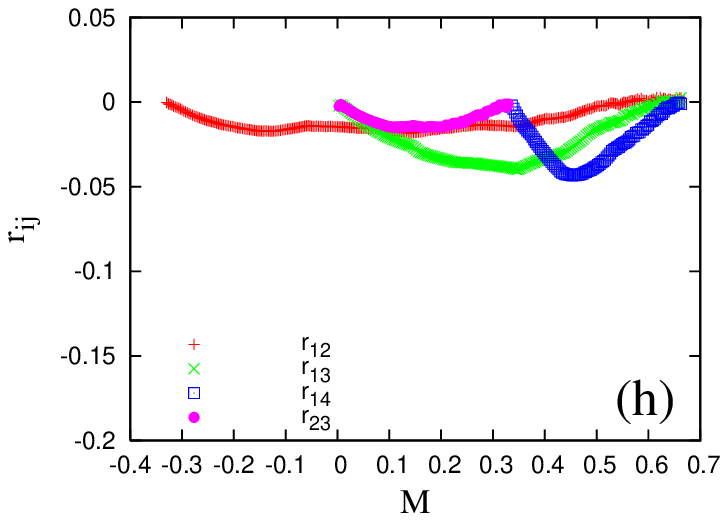}
    \includegraphics[width=2.2in]{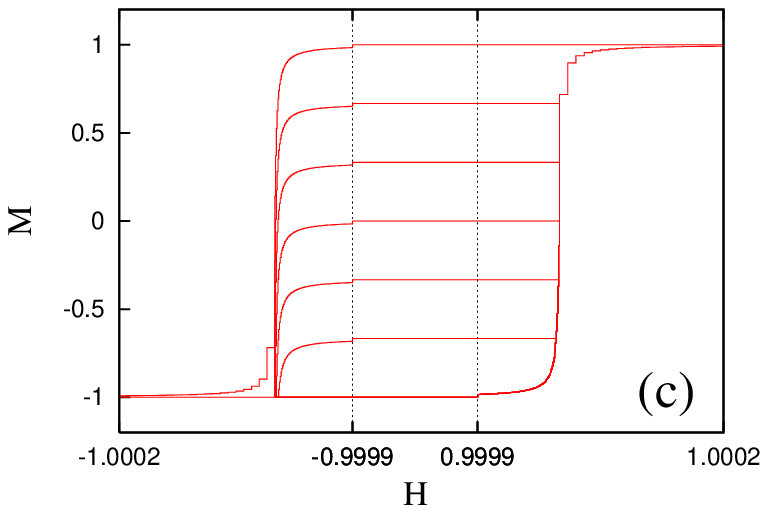}
    \includegraphics[width=2.2in]{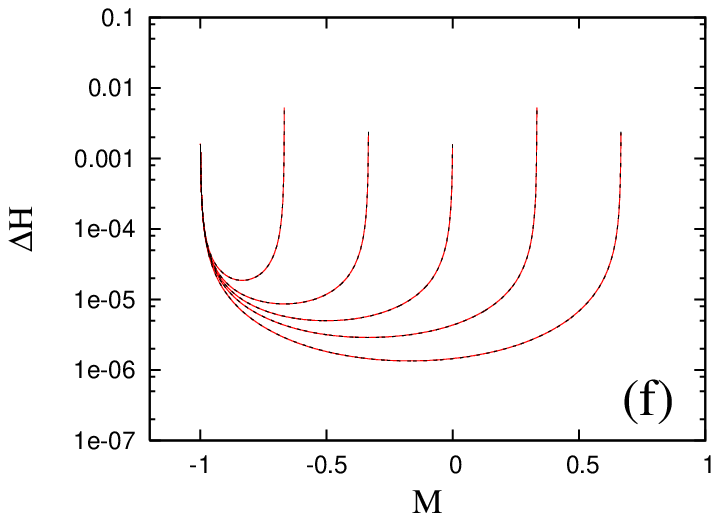}
    \includegraphics[width=2.2in]{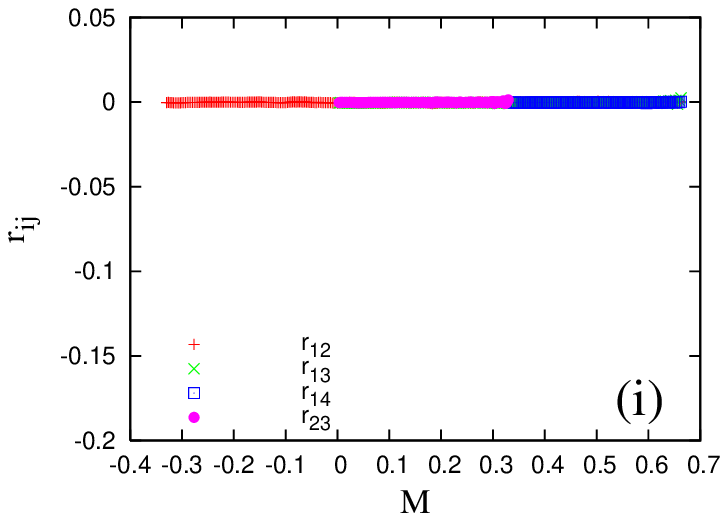}
    \caption{\label{fig:Lorentzian} Numerical results using a Lorentzian distribution
      RSFHM. Rows: (Top, fig.a,d,g): $\sigma=1.6$. (Middle, fig.b.e.h): $\sigma=5$. (Bottom,
      fig.c,f,i): $\sigma=50$. Columns: (Left, fig.a,b,c): $M(H)$ curves, main loop and 5 recoil
      curves. Note that due to the very large $h_0/\sigma$ ratio ($\sim 2\times 10^5$ for
      this system size $1000^2$, chosen to avoid negative $H_\mm{S}$),
      the differences between the major loop and all the recoil curves are
      extremely hard to see from the $M(H)$ plot itself. 
      The small differences will be more clear with log scale as shown in the
      middle column. (Middle, fig.d,e,f): $\Delta H(M, \Delta M)$ curves for
      the 5 recoil curves: (solid lines) numerical result; (dotted lines)
      mean-field approximation. Here, we see that normalized $\Delta H$ values
      are very small compared to the Gaussian $D(H_\mm{S})$ case due to the large $h_0/\sigma$ 
      ratio. (Right, fig.g,h,i): $M$-dependent deviation from redundancy ($r_{ij}(M)$) for all the
      possible recoil curve pairs.}
  \end{figure*}

\begin{figure*}[t]
    \includegraphics[width=2.2in]{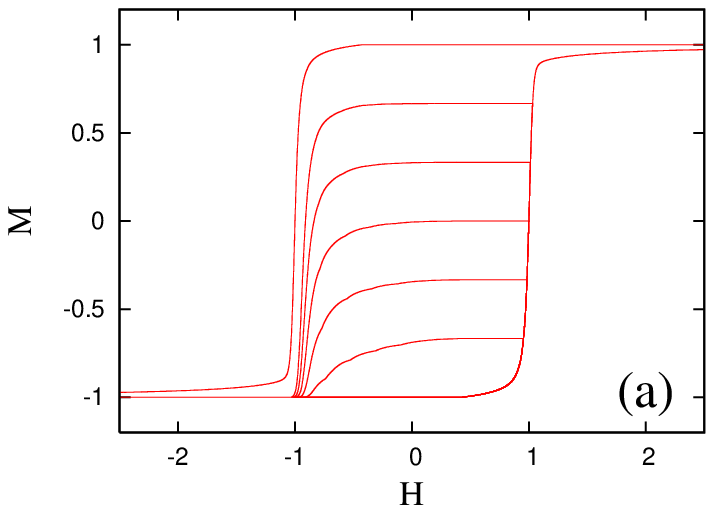}
    \includegraphics[width=2.2in]{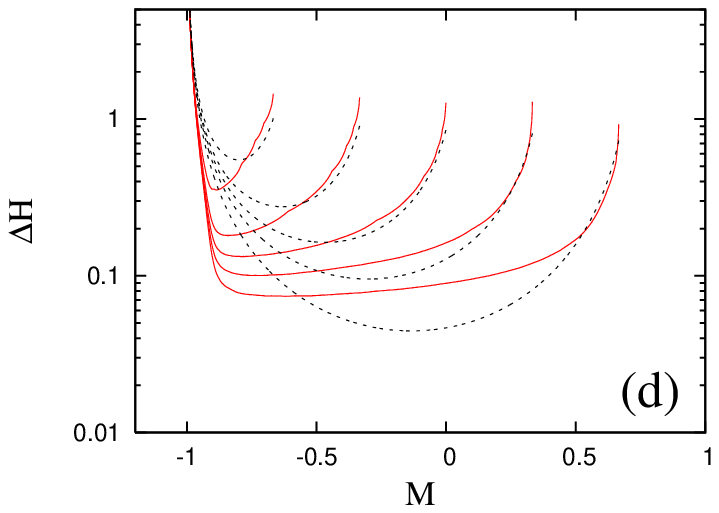}
    \includegraphics[width=2.2in]{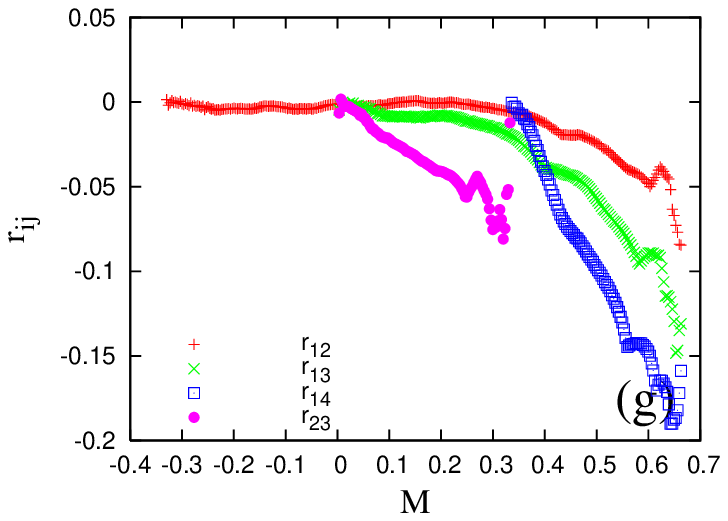}
    \includegraphics[width=2.2in]{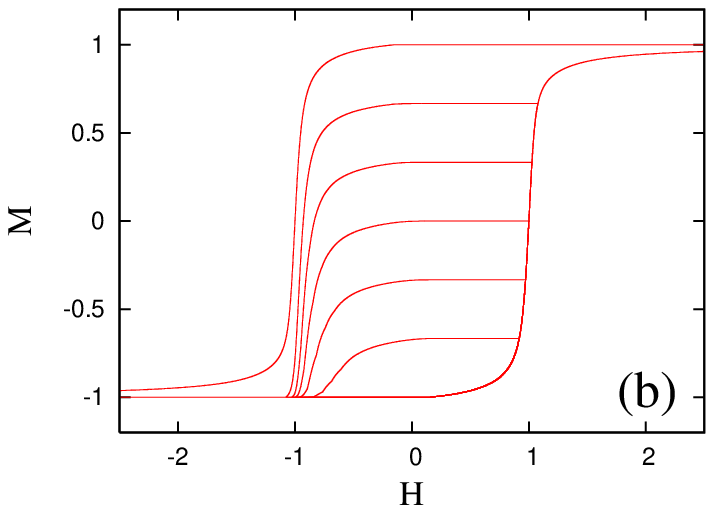}
    \includegraphics[width=2.2in]{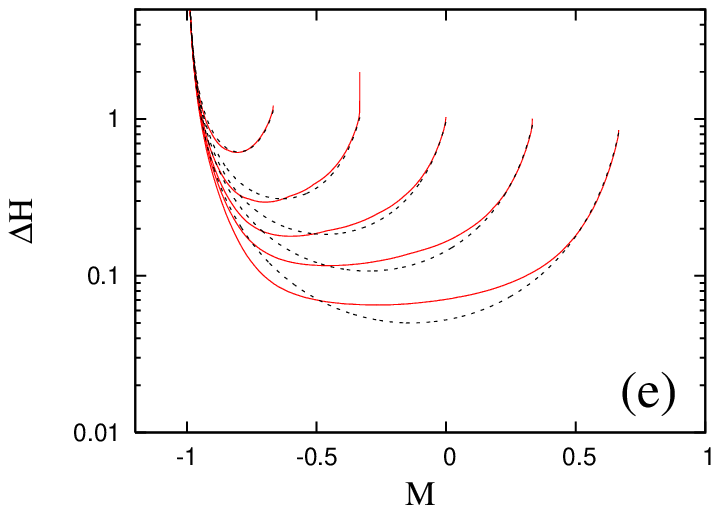}
    \includegraphics[width=2.2in]{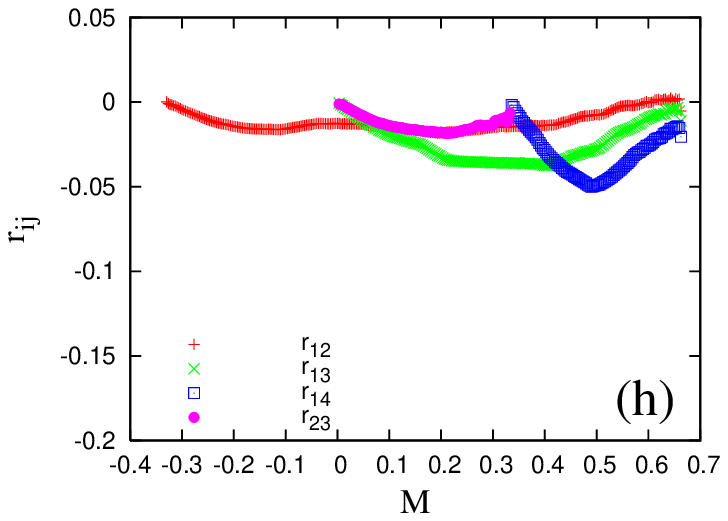}
    \includegraphics[width=2.2in]{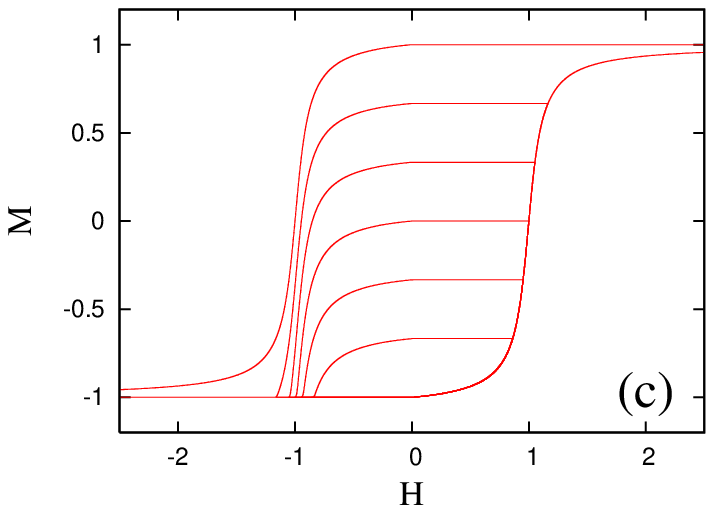}
    \includegraphics[width=2.2in]{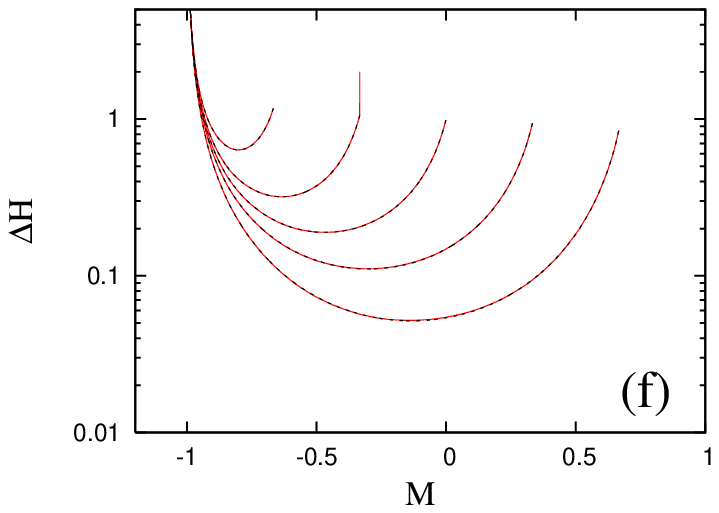}
    \includegraphics[width=2.2in]{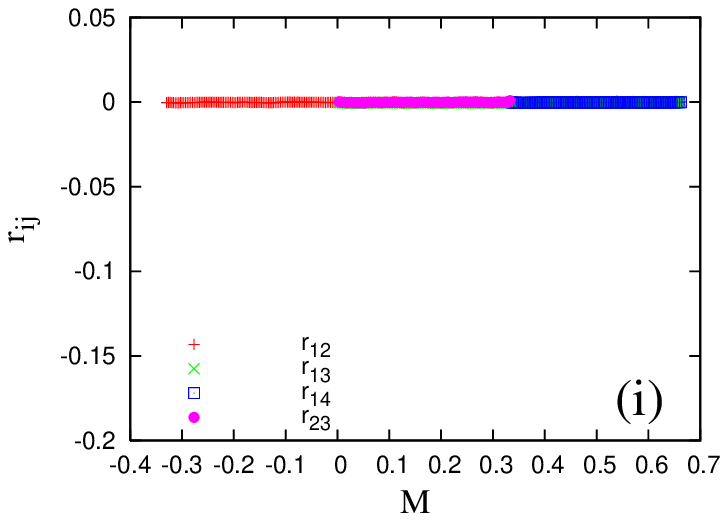}

    \caption{\label{fig:Lorentzian-I} Numerical results using a truncated Lorentzian distribution
      RSFHM. Rows: (Top, fig.a,d,g): $\sigma=1.6$. (Middle, fig.b.e.h): $\sigma=5$. (Bottom,
      fig.c,f,i): $\sigma=50$. Columns: (Left, fig.a,b,c): $M(H)$ curves, main loop and 5 recoil
      curves. (Middle, fig.d,e,f): $\Delta H(M, \Delta M)$ curves for the 5 recoil curves: (solid
      lines) numerical result; (dotted lines) mean-field approximation.
      (Right, fig.g,h,i): $M$-dependent deviation from redundancy ($r_{ij}(M)$) for all the
      possible recoil curve pairs.}
  \end{figure*}

\begin{figure*}[t]
    \includegraphics[width=2.2in]{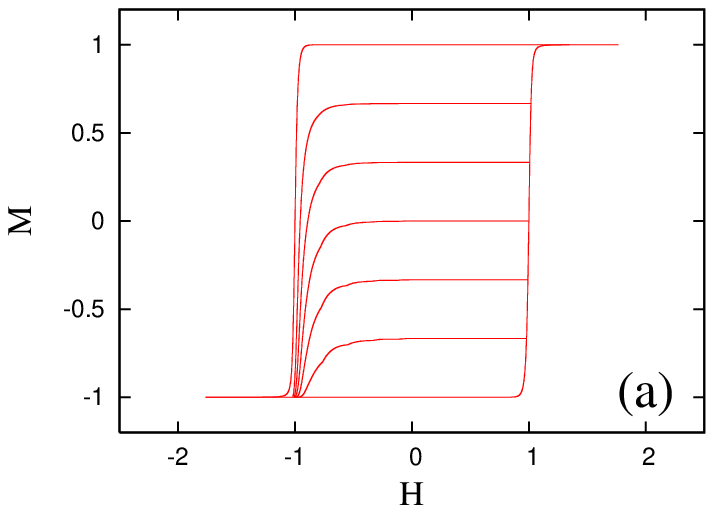}
    \includegraphics[width=2.2in]{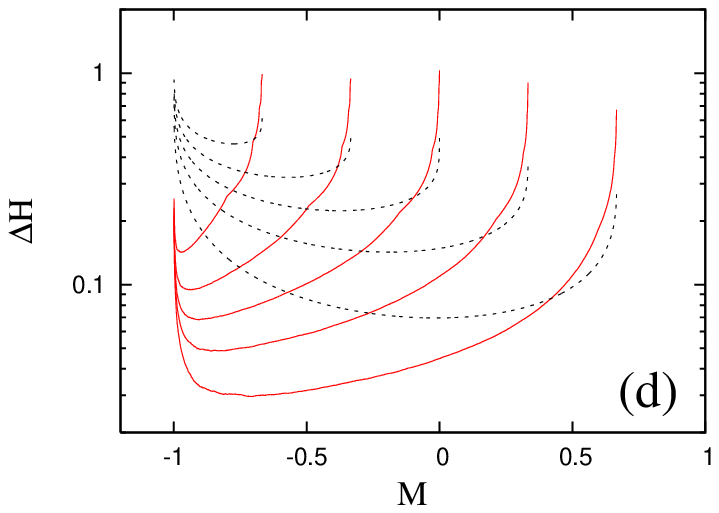}
    \includegraphics[width=2.2in]{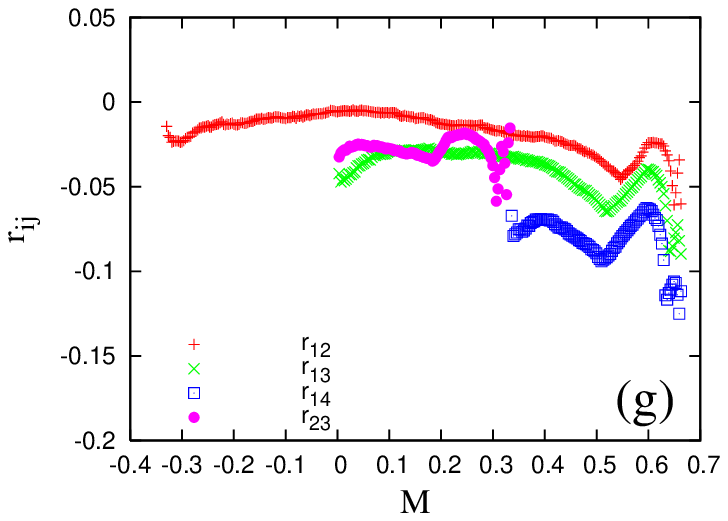}
    \includegraphics[width=2.2in]{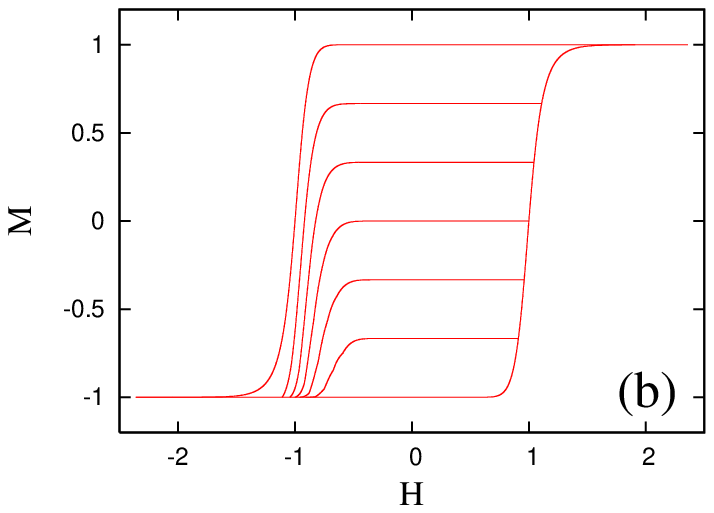}
    \includegraphics[width=2.2in]{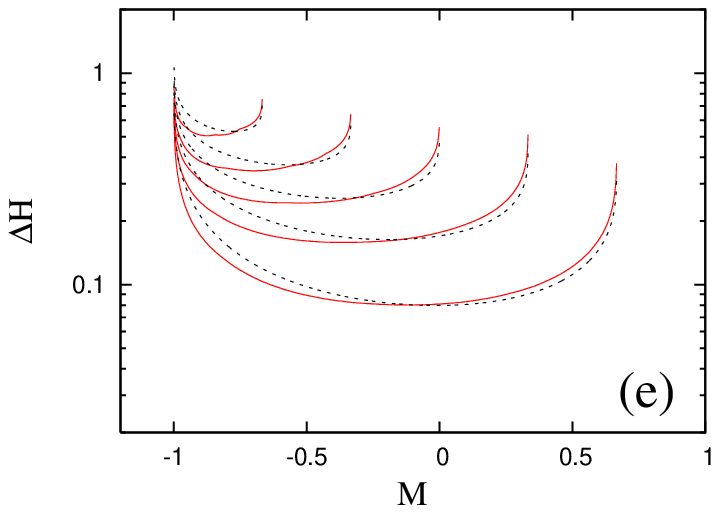}
    \includegraphics[width=2.2in]{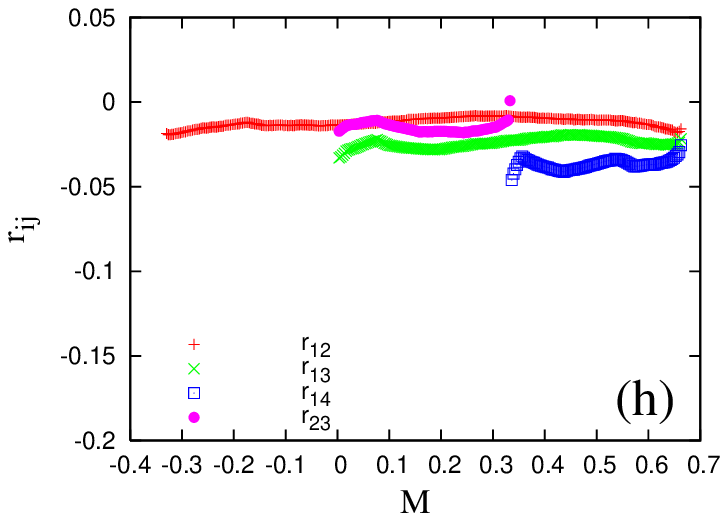}
    \includegraphics[width=2.2in]{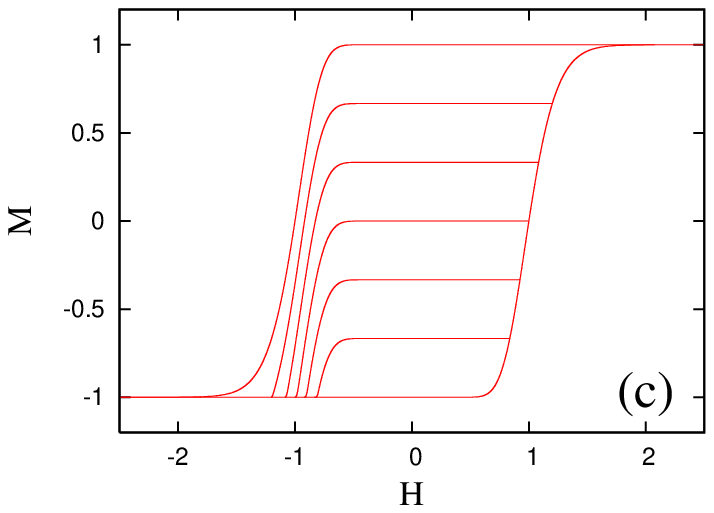}
    \includegraphics[width=2.2in]{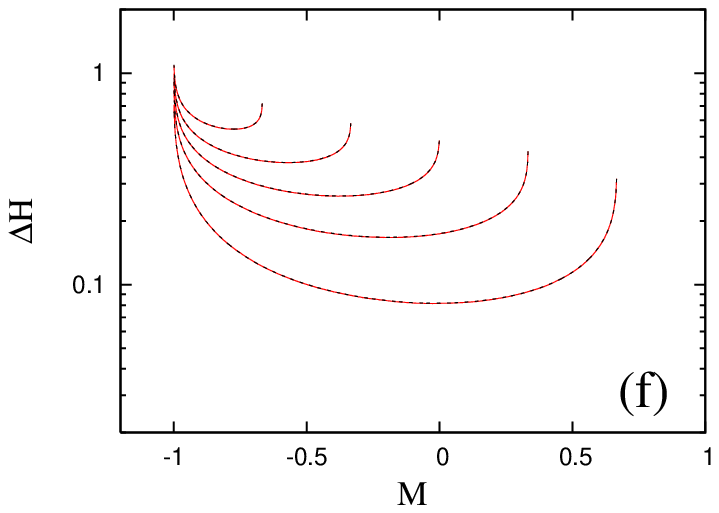}
    \includegraphics[width=2.2in]{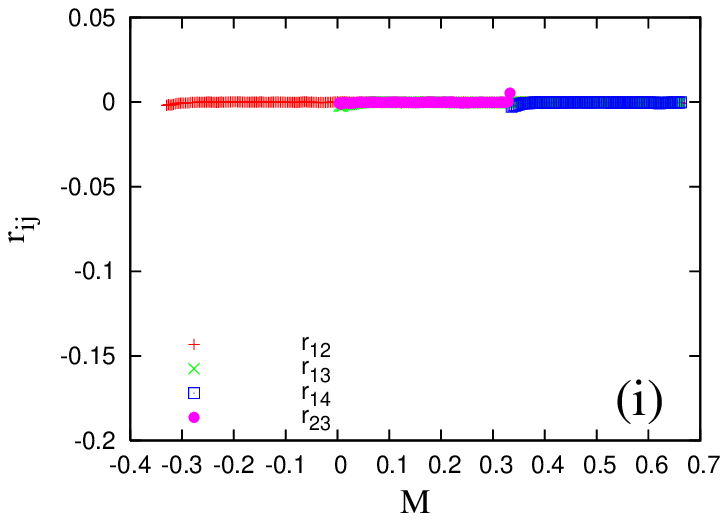}

    \caption{\label{fig:Lognormal} Numerical results using a Lognormal distribution
      RSFHM. Rows: (Top, fig.a,d,g): $\sigma=1.6$. (Middle, fig.b.e.h): $\sigma=5$. (Bottom,
      fig.c,f,i): $\sigma=50$. Columns: (Left, fig.a,b,c): $M(H)$ curves, main loop and 5 recoil
      curves. (Middle, fig.d,e,f): $\Delta H(M, \Delta M)$ curves for the 5 recoil curves: (solid
      lines) numerical result; (dotted lines) mean-field approximation.
      (Right, fig.g,h,i): $M$-dependent deviation from redundancy ($r_{ij}(M)$) for all the
      possible recoil curve pairs.}
  \end{figure*}

  Key results of our numerical hysteresis loop calculations for all these
  different switching field distributions are shown in
  Figs.~\ref{fig:Gaussian}-~\ref{fig:Lognormal}. In each case, we show plots for $\sigma=1.6, 5$
  and 50 only to illustrate the general trends. All the calculations shown here are done
  in $D=2$ dimensions for linear system size $L=1000$, i.e. $L^2=10^6$
  hysterons.

  At first, we discuss the results for the Gaussian $D(H_\mm{S})$ distribution in
  detail. Fig.~\ref{fig:Gaussian} displays the results for different $\sigma$'s
  in different rows: (Top) $\sigma=1.6$. (Middle) $\sigma=5$. (Bottom)
  $\sigma=50$. For each $\sigma$, we calculate a complete set of $M(H)$ curves,
  both the saturation hysteresis loop and recoil curves, as shown in the left
  column of Fig.~\ref{fig:Gaussian}. Note that here and throughout the paper,
  $M$ (or $\Delta M$) is normalized to the saturation value $M_\mm{S}$ and
  $H$ (or $\Delta H$) is normalized to the coercive field
  $H_\mm{C}$. In particular, we choose 5 equally-spaced
  recoil curves, for which the distance to saturation is given by $\Delta M_i
  = i/3$. From the left column of Fig.~\ref{fig:Gaussian}, one can see that the
  hysteresis curves get broader for higher $\sigma$. Note that it is not the
  larger $\sigma$ itself that causes this effect, because this type of
  broadening is taken out due to the normalization of $H$ with $H_\mm{C}$, and
  the constant $h_0/\sigma$ ratio~\cite{note:h0Hc}. The difference in shape
  here actually reflects the fact that for lower $\sigma$, one gets correlated
  magnetization reversal which sharpens the macroscopic switching field
  distribution substantially.

  In the middle column of Fig.~\ref{fig:Gaussian}, we show the corresponding
  $\Delta H(M,\Delta M)$ curves (solid lines) derived from the simulated
  $M(H)$ curves, as well as the mean-field approximation of the $\Delta H(M,
  \Delta M)$ curves (dotted lines) calculated from
  Eq.~\ref{eq:DeltaH-Gaussian}.  The mean-field curves are calculated by
  using the exact input parameter and are not a least square fit. This allows
  for a clear illustration of the deviations from mean-field behavior. 
  From Fig.~\ref{fig:Gaussian}(d), we see that
  for small $\sigma$ (corresponding to strong hysteron interactions) the
  difference between the numerical result and the mean-field approximation is
  large. For intermediate $\sigma$ (corresponding to intermediate hysteron
  interactions), the difference diminishes but is still visible, especially
  near the negative saturation $M=-1$, as shown in
  Fig.~\ref{fig:Gaussian}(e). For high $\sigma$ (corresponding to
  weak hysteron interactions), the difference is so small that it is not
  visible in Fig.~\ref{fig:Gaussian}(f). It should be mentioned that due to
  the constant $h_0/\sigma$ ratio and the normalization of $\Delta H$, the
  mean-field $\Delta H(M, \Delta M)$ curves look almost identical for
  different $\sigma$'s~\cite{note:h0Hc}. Furthermore, it is apparent that the
  $\Delta H(M,\Delta M)$ curves obtained from numerical simulations are
  asymmetric, in particular for small $\sigma$. They show much larger
  deviations from the mean-field approximation on the left hand side,
  i.e. near negative saturation $M=-1$. This can also be seen in the
  hysteresis loops themselves, where the curves seem to bundle up near
  negative saturation for small $\sigma$ values.

  In the right column of Fig.~\ref{fig:Gaussian}, we show  numerical values for
  the deviation from redundancy measures
  $r_{ij}(M)$ which are calculated from the simulated recoil curves shown in
  the left column. Due to the definition range of $r_{ij}(M)$, only the
  recoil curve pairs $(i,j)= (1,2)$, $(1,3)$, $(1,4)$ and
  $(2,3)$ produce data. From Fig.~\ref{fig:Gaussian}(g), we see that for small
  $\sigma$ the deviation from data redundancy is quite substantial for the
  whole $M$ definition range and for all the recoil curve pairs. For intermediate
  $\sigma$, the deviation becomes smaller but is still visible, as shown in
  Fig.~\ref{fig:Gaussian}(h). For high $\sigma$, the deviation is almost
  negligible in the whole $M$ definition range and for all recoil curve pairs,
  as one can see from Fig.~\ref{fig:Gaussian}(i). Small non-vanishing values occurring
  at the boundaries of the $M$ definition range are just due to
  the incompletely canceled finite size inaccuracies, as discussed in Sec.~\ref{sec:r}.

  For the other $D(H_\mm{S})$ distributions, we observe very similar
  results as shown in Figs.~\ref{fig:Lorentzian}--\ref{fig:Lognormal}.
  Thus, one has to realize that for small $\sigma$, i.e. strong hysteron
  interactions, the $\Delta H(M,\Delta M)$ method is not accurate.
  The mean-field approximation doesn't match the numerical result and
  deviations from redundancy are large. This is the expected result because
  once coupling dominates the magnetization reversal the mean-field approximation
  will not be valid any more. On the other hand, for large $\sigma$, i.e.
  weak hysteron interactions, the $\Delta H(M,\Delta M)$ method works very well, which is
  indicated by both the small deviation from redundancy and the match of the
  mean-field approximation to the numerical results.

  \subsection{Emergent feature of the $\Delta H(M,\Delta M)$ method}

\begin{figure}[t]
    \includegraphics[width=0.5\textwidth]{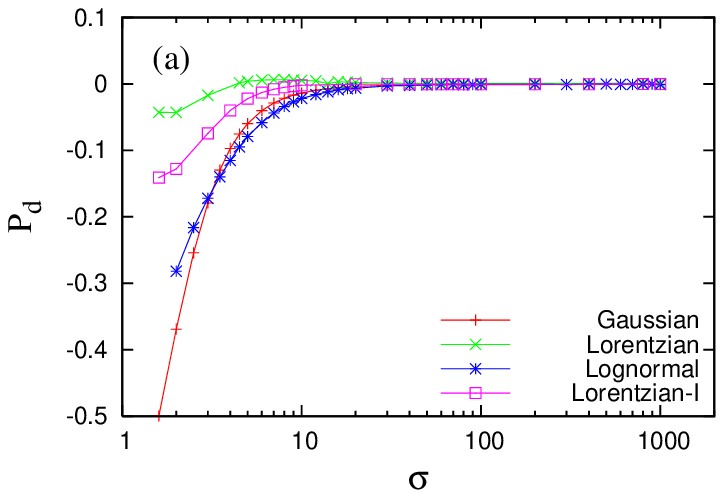}
    \includegraphics[width=0.5\textwidth]{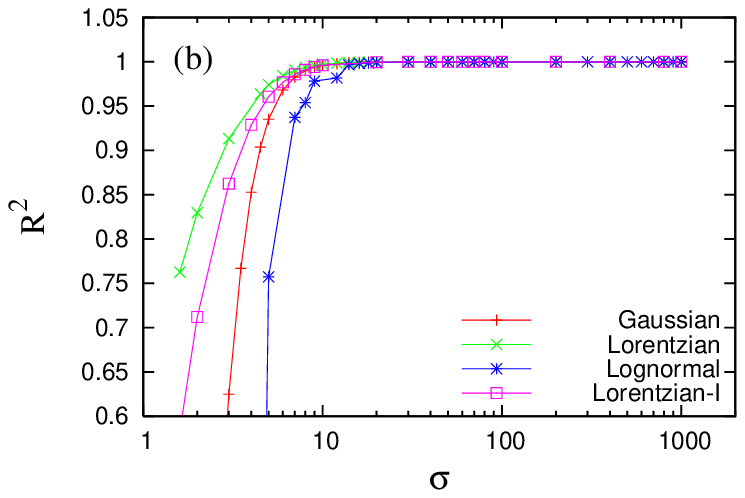}
    \includegraphics[width=0.5\textwidth]{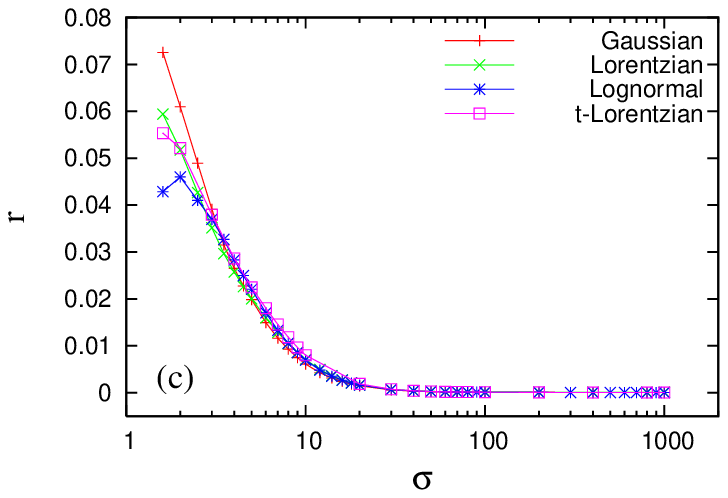}
    \caption{\label{fig:alldistributions} Reliability measures for the
      different $D(H_\mm{S})$: (a) $P_\mm{d}$. (b) $R^2$. (c) $r$.}
  \end{figure}

  The similarities in the failure of the $\Delta H(M, \Delta M)$
  method for different $D(H_\mm{S})$ distributions indicates that this
  method may be not very sensitive to the particular type of distribution in
  general. To study this further, we plot the reliability measures against the
  tuning parameter $\sigma$ for all $D(H_\mm{S})$ distributions in
  Fig.~\ref{fig:alldistributions}.

  The fit quality measures $P_\mm{d}$ and $R^2$ are shown in
  Fig.~\ref{fig:alldistributions}(a) and (b), respectively.
  We see that $P_\mm{d}$ approaches 0 with increasing $\sigma$, which means
  that for high $\sigma$, the input value of $\sigma$ can be recovered with
  very high accuracy by fitting the $\Delta H(M,\Delta M)$ data to
  the mean-field fit function. In other
  words, the $\Delta H(M,\Delta M)$ method works well for high
  $\sigma$. Furthermore, we see that with increasing $\sigma$, $R^2$
  approaches 1 corroborating a successful fit of the $\Delta H(M,\Delta M)$
  data in this regime. The average deviation from redundancy ($r$) is shown in
  Fig.~\ref{fig:alldistributions}(c). It is clearly seen that with
  increasing $\sigma$, $r$ approaches 0, i.e. data redundancy is obtained,
  which is the key feature of the mean-field approximation.

  The reliability range of the $\Delta H (M, \Delta M)$ method can be
  obtained from the reliability measures shown in
  Fig.~\ref{fig:alldistributions}. We find that for all the four
  $D(H_\mm{S})$ distributions, with the definitions of $\sigma$ given
  in Sec.~\ref{sec:method}, the $\Delta H (M, \Delta M)$ method works
  virtually perfect for $\sigma \ge \sigma_0$ with $\sigma_0$ being
  approximately equal to 20. Here, $\sigma_0$ is just a rough
  criterion, above which the reliability measures have merged into
  their mean-field approximation values. It should be emphasized that
  the $\sigma_0$ values obtained from all three reliability measures
  are fairly consistent. If there are differences at all, $r$ appears
  to show the highest sensitivity to deviations from the mean-field
  approximation, while $R^2$ seems to be slightly less sensitive. This
  is important, because $r$ can be evaluated without any fit from a
  data set alone. So, the independent reliability test is the most
  sensitive measure and gives one confidence that not only the $\Delta H
  (M, \Delta M)$ method fails in a well-defined way, but also that one
  can very reliably check for this failure mode.

  Finally, we note that the different $r$ vs. $\sigma$ curves in Fig.
  ~\ref{fig:alldistributions}(c) for the different distribution types
  track each other almost exactly. However, this particular
  observation dependents somewhat on how exactly we define $\sigma$ in
  the various $D(H_\mm{S})$ distributions, because only for the
  Gaussian and Lognormal distribution are we using the natural
  definition given by the standard deviation. Thus, the curve collapse
  seen in Fig.~\ref{fig:alldistributions}(c) might be partially artificial.
  $P_\mm{d}$ and $R^2$ on the other hand do not exhibit such a
  collapse, not even for the Gaussian and Lognormal distributions as
  is apparent from Fig.~\ref{fig:alldistributions}(a) and (b). The
  fact that the two kinds of reliability measures show different
  behavior can be understood in the following way. As mentioned in the end of
  Sec.~\ref{sec:fitquality} and Sec.~\ref{sec:r}, finite size
  inaccuracies at the definition range boundaries of the recoil curves will
  not affect the calculation of $r$ very much due to the cancellation
  effect. But it will affect the calculation of the fit quality
  measures, both $P_\mm{d}$ and $R^2$. Generally speaking, the shape of the
  hysteresis loops and recoil curves depends on the particular type of the
  chosen $D(H_\mm{S})$. Consequently, the finite size inaccuracies will also
  depend on $D(H_\mm{S})$. As a result, we see different (similar) behaviors of
  the fit quality measures (deviation from redundancy measure) for different
  $D(H_\mm{S})$ distributions at small $\sigma$. In this sense, it is
  natural to choose the deviation from redundancy measure $r$ as the best
  measure to determine the reliability range of the $\Delta H(M, \Delta M)$
  method.

  \section{\label{sec:sum}Summary}

  We study the $\Delta H(M, \Delta M)$ method and its reliability by
  means of numerical simulations of the zero-temperature random
  switching field hysteron model. We present strong evidence that the
  $\Delta H(M, \Delta M)$ method, which is based on the mean-field
  approximation, has a well-defined reliability range. This
  reliability range can be checked with two types of independent
  measures: deviation from redundancy and fit quality. The former is
  the superior tool because it is calculated from the data set alone
  and is independent from any inaccuracies that might be induced by
  data fitting procedures itself.

  \section{Acknowledgment}

  We thank M. Delgado for valuable discussions. Work done at UIUC was
  conducted on the Beowolf cluster of the Materials Computation Center
  at UIUC and acknowledges the support of NSF Grant No. DMR 0314279
  and NSF Grant No. DMR 03-25939 ITR (Materials Computation Center).

  \appendix
  \section{\label{sec:app:mapping} Mapping RSFHM to RFIM}

  For the zero-temperature non-equilibrium RFIM, a local metastable dynamics
  has been introduced by Sethna \emph{et al.}~\cite{Sethna-93} to study the
  disorder-induced phase transition in the hysteretic behavior at $T=0$:
  Initially, all spins point DOWN, as $H$ is slowly increased from
  $-\infty$ to $\infty$ and decreased back to $-\infty$, each spin flips
  deterministically when its effective local field \be h^{\rm eff}_i = J
  \sum_{j} s_j + H + h_i \ee changes sign. For the zero-temperature
  non-equilibrium RSFHM, we introduce a similar local metastable dynamics:
  Initially, all spins point DOWN, as $H$ is slowly increased from $-\infty$
  to $\infty$ and decreased back to $-\infty$, each hysteron flips
  deterministically when its effective local field \be H^{\rm eff}_i = J
  \sum_{j} S_j + H + \text{sgn}(S_i) \ {H_\mm{S}}_i\ee changes sign.

  Considering the only slight difference between the RFIM and the RSFHM, it is possible
  to introduce a simple mapping: $h_i \leftrightarrow \text{sgn}(S_i) \
  {H_\mm{S}}_i$ with ${H_\mm{S}}_i > 0$. This enables us to calculate the
  $M(H)$ curve of the interacting hysterons by means of simple software adaptation.

  \section{\label{sec:app:proof_redundancy} Proof of Data Redundancy in Mean-Field theory}
  As shown in Fig.~\ref{fig:redundancy_MH}, we choose six states (A, B, P, Q,
  U and V) from the major loop and recoil curves with 

  \begin{figure}[t]
    \includegraphics[width=0.5\textwidth]   {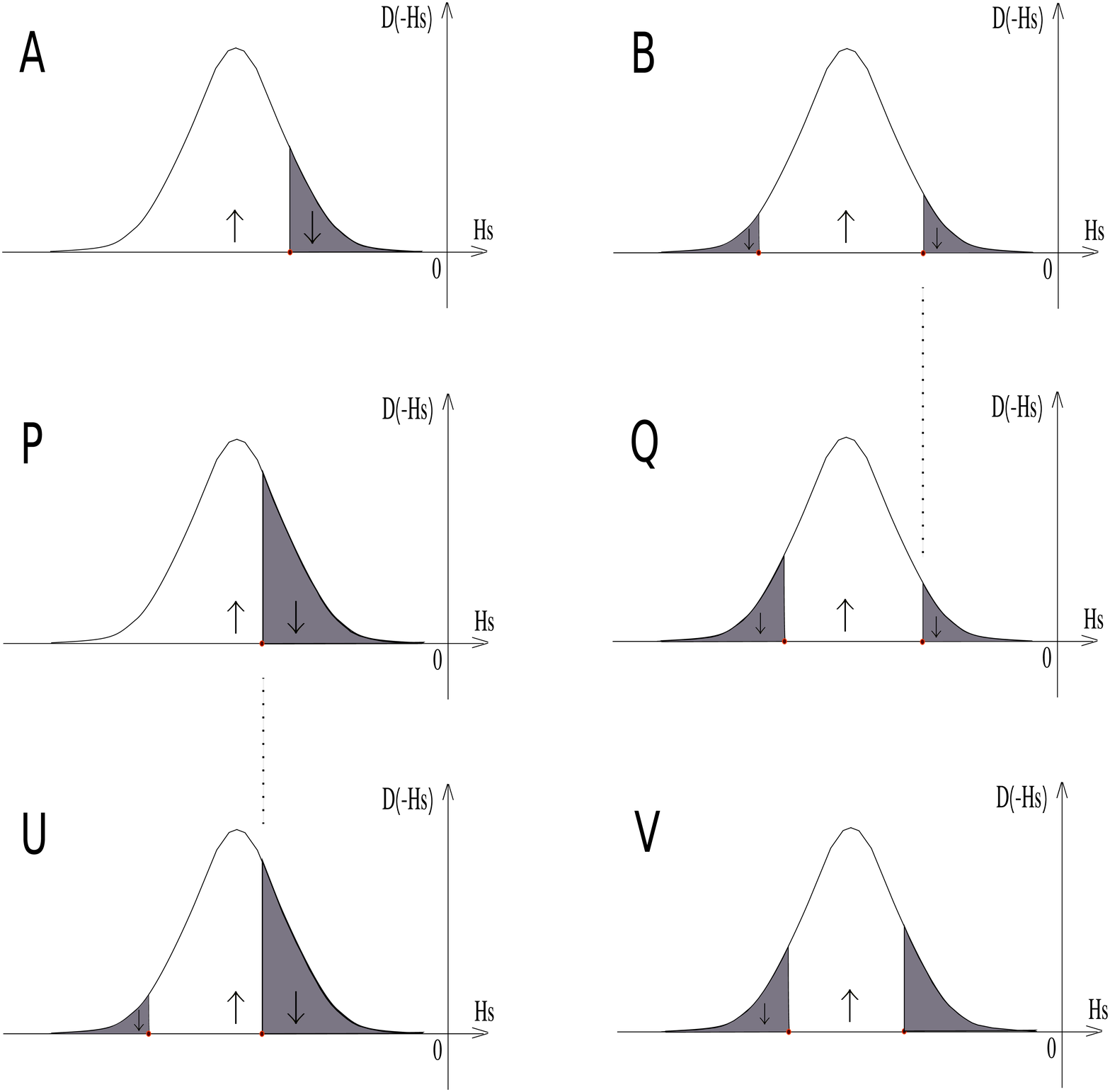}
    \caption{\label{fig:redundancy_states} Hysteron distributions for the
      various states (A, B, P, Q, U and V) shown in
      Fig.~\ref{fig:redundancy_MH}. The shadowed areas symbolize
      DOWN-hysterons ($S_i=-1$), the open areas correspondingly UP-hysterons
      ($S_i=+1$).} 
  \end{figure}

  \begin{subequations}
    \begin{align}
      M_\mm{A}&= M_\mm{B} =M \\
      M_\mm{P}&= M_\mm{Q} =M-\Delta M_{j}+\Delta M_{i} \\
      M_\mm{U}&= M_\mm{V} =M-\Delta M_{j}
    \end{align}
  \end{subequations} and $M$ is an arbitrary value within $[-1,1-\Delta
    M_i]$. The hysteron distributions for the six states are shown in
  Fig.~\ref{fig:redundancy_states}. Shadowed areas denote DOWN-hysterons
  ($S_i=-1$), while open areas indicate UP-hysterons ($S_i=+1$). In the following,
  we use the symbol $n_{\downarrow}$ and $n_{\uparrow}$ for the density of DOWN-hysteron and
  UP-hysteron, respectively. For the states chosen from the upper branch of the
  major loop with magnetization $M_0$, it is easy to get
  $n_{\downarrow}=(1-M_0)/2$. For example,

  \begin{subequations}
    \begin{align}
      n_{\downarrow}(A)&=(1-M)/2 \\
      n_{\downarrow}(P)&=(1-M-\Delta M_i + \Delta M_j)/2
    \end{align}
  \end{subequations}

  For states chosen from the recoil curves with
  magnetization $M_0$, there are two shadowed areas, which can be denoted as
  $n_{\downarrow 1}$ and $n_{\downarrow 2}$. Note that the left one
  $n_{\downarrow 1}$ is just due to the distance from saturation at the
  starting point of the recoil curve, i.e. $\Delta M$, so that $n_{\downarrow 1}=
  \Delta M/2$. For example, B and U are picked from the $i$-th recoil curve
  with distance from saturation $\Delta M_i$, so that 

  \be n_{\downarrow 1}(B) = n_{\downarrow 1}(U) = \Delta M_i/2 \ee

  Similarly, for Q and V, we have 
  \be n_{\downarrow 1}(Q) = n_{\downarrow 1}(V) = \Delta M_j/2 \ee

  Combining this with the number conservation equation
  $n_{\downarrow}=n_{\downarrow 1} + n_{\downarrow
    2}=(1-M_0)/2$, we find

  \begin{subequations}
    \begin{align}
      n_{\downarrow 2}(B)&=(1-M-\Delta M_i)/2 \\
      n_{\downarrow 2}(Q)&=(1-M-\Delta M_i)/2 \\
      n_{\downarrow 2}(U)&=(1-M-\Delta M_i+ \Delta M_j)/2 \\
      n_{\downarrow 2}(V)&=(1-M)/2
    \end{align}
  \end{subequations}

  It follows that

  \begin{subequations}
    \begin{align}
      n_{\downarrow}(A) &= n_{\downarrow 2}(V) \\
      n_{\downarrow}(P) &= n_{\downarrow 2}(U) \\
      n_{\downarrow 2}(B) &= n_{\downarrow 2}(Q)
    \end{align}
  \end{subequations}

  and

  \bw
  \begin{subequations}
    \begin{align}
      H_\mm{A}+H_\mm{int}(M)&=H_\mm{V}+H_\mm{int}(M-\Delta M_j) \\
      H_\mm{P}+H_\mm{int}(M-\Delta M_j + \Delta M_i)&=H_\mm{U}+H_\mm{int}(M-\Delta M_j) \\
      H_\mm{B}+H_\mm{int}(M)&=H_\mm{Q}+H_\mm{int}(M-\Delta M_j + \Delta M_i)
    \end{align}
  \end{subequations}

  So, overall we find

  \be \left( H_\mm{B} - H_\mm{A} \right) + \left( H_\mm{V} - H_\mm{U} \right)
  - \left( H_\mm{Q} - H_\mm{P} \right) = 0 \ee
  Q.E.D
  \ew

  \section{\label{sec:app:truncate_Lorentzian} $\Delta(H, \Delta M)$ formula
    for the truncated Lorentzian Distribution}
  In numerical simulation, the random switching fields with any distributions are
  generated by a random number generator. To avoid any negative tails in the
  Lorentzian distribution, we can artificially suppress any negative random
  numbers and instead create another random number for the switching field
  until we get a positive one. The corresponding switching field distribution
  is then represented by the truncated Lorentzian distribution $D_\mm{L_t}(H_\mm{S})$:

  \be
  D_\mm{L_t}(H_\mm{S}) = \left \{ \begin{array}{ll}
    0   & \textrm{ for } H_\mm{S}<0 \\
    \frac{2w}{\pi}\frac{C}{w^2+4(H_\mm{S}-h_0)^2} & \textrm{ for }
    H_\mm{S} \ge 0
  \end{array} \right.
  \ee

  with $C$ given by the condition that $\int_{-\infty}^{+\infty} D_\mm{L_t}(H_\mm{S}) \ \ud H_\mm{S} =
  1 $. Specifically, one finds that
  \be C = \left( \frac{1}{2}+\frac{\theta}{\pi} \right )^{-1} \ee
  with $\theta = \tan^{-1}(\frac{2h_0}{w})$.

  For the truncated Lorentzian distribution, we define the disorder parameter
  $\sigma = w$ to make it comparable with the Lorentzian distribution.

  The integral function is given by:

  \bea
  I_\mm{L_t}(x)
  &=&  \int_{-\infty}^{x} D_\mm{L_t}(H_\mm{S}) \ \ud H_\mm{S} \nn
  &=&  \frac{C}{\pi} \left[ \tan^{-1}\left(\frac{2(x-h_0)}{w}\right) + \theta \right]
  \eea
  so that \be I^{-1}_\mm{L_t}(y) = h_0 + \frac{w}{2} \
  \tan\left(\frac{\pi}{C}y - \theta\right) \ee 

  From this result, we derive 

  \bea
  \Delta H_\mm{L_t}(M,\Delta M)
  &=& \frac{w}{2} \Big\{ \tan \big[ (\frac{\pi}{2} + \theta) \left(\frac{1-M}{2}\right) - \theta \big] \nn
  & & -  \tan \big[ (\frac{\pi}{2} + \theta) \big(\frac{1-M-\Delta M}{2}\big) - \theta \big] \Big\} \nn
  & &  \eea
  as the analytic mean-field solution. We notice that due to the
  truncation induced asymmetry in the distribution itself, $\Delta
  H_\mm{L_t}(M,\Delta M)$ depends on both $w$ and
  $h_0$.


\begin{thebibliography}{99}
\bibitem{note:rij} \bibinfo{note}{Since the definition range of $M$ in
    $r_{ij}(M)$ is $[\Delta M_j -1, 1- \Delta M_i]$, we must have $1- \Delta M_i
    > \Delta M_j -1$, i.e. $\Delta M_i + \Delta M_j < 2$.}
\bibitem{note:h0Hc} \bibinfo{note}{This statement only makes sense if
    $H_\mm{C}$ and $h_0$ are very similar, which should be fulfilled as long
    as the hysteron coupling is not too strong and the main cause of
    hysteresis is the single hysteron hysteresis. In fact, in our simulations,
    we find that $h_0$ and $H_\mm{C}$ are almost identical (or very similar)
    for high $\sigma$. Moreover, one could make this an exact statement by
    normalizing $H$ and $\Delta H$ to $h_0$.}
\end{thebibliography}

\end{document}